\documentclass[preprint,12pt]{elsarticle}
\biboptions{sort&compress}

\usepackage{amsmath,amssymb,amsfonts}

\usepackage{graphicx}
\usepackage{subcaption} 

\usepackage{booktabs}   
\usepackage{makecell}   
\usepackage{multirow}
\usepackage{adjustbox}  

\usepackage{xltabular}    
\keepXColumns
\renewcommand{\arraystretch}{1.1}

\usepackage{siunitx}
\usepackage{algorithm}
\usepackage{algorithmic}

\usepackage{xcolor}
\usepackage{textcomp}
\usepackage{soul}
\usepackage[normalem]{ulem} 

\usepackage{hyperref}

\renewcommand{\arraystretch}{1.1}

\newcommand{\INPUT}{\REQUIRE}
\newcommand{\OUTPUT}{\ENSURE}

\newcommand{\ie}{{\it i.e.}}
\newcommand{\eg}{{\it e.g.}}

\def\BibTeX{{\rm B\kern-.05em{\sc i\kern-.025em b}\kern-.08em
    T\kern-.1667em\lower.7ex\hbox{E}\kern-.125emX}}
\newcommand{\compressformat}{CSSC}

\journal{Information Science}

\begin{document}

\begin{frontmatter}

\title{Efficient Privacy-Preserving Sparse Matrix-Vector Multiplication Using Homomorphic Encryption
}

\author[1]{Yang Gao*}
\ead{yang.gao@ucf.edu}

\author[2]{Gang Quan}
\ead{gaquan@fiu.edu}

\author[3]{Wujie Wen}
\ead{wwen2@ncsu.edu}

\author[1]{Scott Piersall}
\ead{spiersall@ucf.edu}

\author[1]{Qian Lou}
\ead{qian.lou@ucf.edu}

\author[1]{Liqiang Wang}
\ead{Liqiang.Wang@ucf.edu}

\affiliation[1]{
  organization={University of Central Florida},
  city={Orlando},
  state={FL},
  country={USA}
}

\affiliation[2]{
  organization={Florida International University},
  city={Miami},
  state={FL},
  country={USA}
}

\affiliation[3]{
  organization={North Carolina State University},
  city={Raleigh},
  state={NC},
  country={USA}
}

\cortext[cor1]{Corresponding author}

\begin{abstract}
Sparse matrix–vector multiplication (SpMV) is a fundamental operation in scientific computing, data analysis, and machine learning. When the data being processed are sensitive, preserving privacy becomes critical, and homomorphic encryption (HE) has emerged as a leading approach for addressing this challenge. Although HE enables privacy-preserving computation, its application to SpMV has remained largely unaddressed. 
To the best of our knowledge, this paper presents the first framework that efficiently integrates HE with SpMV, addressing the dual challenges of computational efficiency and data privacy. In particular,
we introduce a novel compressed matrix format, named Compressed Sparse Sorted Column (CSSC), which is specifically designed to optimize encrypted sparse matrix computations. By preserving sparsity and enabling efficient ciphertext packing, CSSC significantly reduces storage and computational overhead. 
Our experimental results on real-world datasets demonstrate that the proposed method achieves significant gains in both processing time and memory usage. 
This study advances privacy-preserving SpMV and lays the groundwork  for secure applications in federated learning, encrypted databases, and scientific computing, and beyond.
\end{abstract}


\end{frontmatter}

\section{Introduction}
Sparse matrix-vector multiplication (SpMV) is a core computational primitive that underpins a wide range of applications, such as scientific simulations, machine learning, optimization, and graph analysis. However, in many of these scenarios, the input data {are} inherently sensitive, such as medical records, financial transactions, user behavior logs, or proprietary simulation data. Computation performed on such data may introduce serious privacy vulnerabilities if not protected. Even indirect exposure to intermediate results can lead to unintended leakage of confidential information.

Homomorphic Encryption (HE) was first introduced by Rivest, Adleman, and Dertouzos
to enable computation over encrypted data while preserving privacy~\cite{rivest1978data},
and was later realized in a fully homomorphic form by Gentry using ideal lattices~\cite{gentry2009fully}.
As a result, HE has emerged as an effective tool for addressing security and privacy issues
in outsourced and untrusted computing environments.
 HE ensures data confidentiality during transit and processing by guaranteeing that the decrypted results are identical to the outcome when the same operations are applied to the data in plaintext. 
While HE offers a promising foundation for secure computation, current HE-based matrix-vector multiplication techniques are designed almost exclusively for dense matrices. They ignore sparsity, leading to devastating effectiveness {degradation} when applied to sparse matrices. 

Classical plaintext-domain sparse formats such as CSR/CSC/COO are fundamentally mismatched with homomorphic evaluation.
They perform unnecessary operations on zero elements, inflate storage costs, and drastically slow down execution. Moreover, these dense approaches often focus on improving ciphertext slot utilization {without accounting for the irregular structure of sparse data,} leading to significant overhead. Therefore, specific homomorphic encryption methods for SpMV are fundamentally lacking and need to be developed.
Prior studies on HE mainly focus on reducing general HE latency through different techniques such as GPU-accelerated attribute-based encryption~\citep{gpu_abe_tpds2023}, encrypted cloud offloading frameworks~\citep{fhe_cloud_tpds2023}, or  parallel processing using Single Instruction Multiple Data (SIMD) \citep{gao2024secure}{. T}hey do not address the unique performance challenges that arise when applying HE to sparse data. In contrast, our work targets this underexplored setting and proposes tailored techniques to improve efficiency.

We present, to our knowledge, the \emph{first} framework that supports SpMV under HE by introducing a new HE-aware sparse format, \emph{Compressed Sparse Sorted Column (CSSC)}, together with an end-to-end packing and aggregation pipeline. CSSC reorders and aligns non-zero elements so that each non-zero–vector pair maps to a corresponding ciphertext slot, enabling exactly one HE Multiplication between ciphertexts and one HE Constant Multiplication between ciphertext and plaintext per aligned pair and minimizing rotations and additions during reduction through chunked, binary-tree accumulation and lightweight masking. Building on this aligned structure, our framework further incorporates ciphertext chunking, vector reorganization, and optimized batched multiplications and additions, collectively yielding {substantial} empirical speedups and substantial memory savings across diverse real-world sparse matrices.

Extensive experiments on a diverse set of real-world sparse matrices demonstrate that our method consistently achieves substantial performance improvements over existing methods. On average, it yields over 100× speedup in runtime and more than 5× reduction in memory usage, with peak improvements reaching up to three orders of magnitude. These results validate the effectiveness of our design and algorithmic optimizations under homomorphic encryption.

Our contributions are summarized as follows:
\begin{itemize}
    \item \textbf{The first encrypted sparse SpMV with both operands encrypted.}
    We present, to our knowledge, the first framework that performs sparse {matrix-vector} multiplication where both matrix and vector are encrypted under homomorphic encryption, enabling end to end confidentiality.

    \item \textbf{An HE-aware sparse format that resolves the existing sparse matrix formats' mismatch.}
    CSR, CSC, and COO are {ill-suited} for fully homomorphic encryption as irregular sparsity triggers costly slot realignment, rotations, masking, and irregular reductions. We introduce Compressed Sparse Sorted Column (CSSC), an HE-aware sparse format that packs non-zeros to align with vector slots, thereby reducing rotations and additions during aggregation while preserving sparsity for compact storage.

    \item \textbf{A low depth, batched encrypted pipeline.}
    We develop an {end-to-end} pipeline that includes ciphertext chunking, vector reorganization, deterministic slot alignment, and logarithmic depth aggregation. The pipeline executes exactly one homomorphic multiplication per aligned pair and a small number of rotations and additions per chunk, which keeps the multiplicative depth at one and amortizes costs through batching.

\end{itemize}

\section{Background and Related Work}

This section {reviews} background and related work: (1) Homomorphic encryption and its core operations; (2) Matrix-vector multiplication under HE, including both dense and sparse settings; and (3) traditional sparse matrix-vector multiplication (SpMV) in plaintext and its incompatibility with HE.

\subsection{Homomorphic Encryption}
\label{sec:he}
Homomorphic encryption (HE) allows computations to be performed directly on encrypted data. {Modern lattice-based schemes such as BFV, originally proposed by Fan and
Vercauteren~\cite{fan2012BFV} and later refined by Brakerski~\cite{Bfv12},
as well as BGV~\cite{brakerski2014BGV} and CKKS~\cite{cheon2017CKKS},}support encrypted arithmetic over vectors.

Most practical HE computations rely on \emph{SIMD packing}~\citep{smart2014fully}, which encodes multiple plaintext values into one ciphertext, and enables {a} single HE operation to be performed on all data elements in the ciphertext simultaneously.

As a representative example, the BFV scheme supports four main encrypted operations: Addition, Ciphertext-Ciphertext Multiplication, Ciphertext-Constant
Multiplication, and Rotation, as shown below, where ciphertexts $ct_x = Enc(x_0, x_1, ..., x_L)$, $ct_y = Enc(y_0, y_1, ..., y_L)$, and a plaintext $pt = (p_0, p_1, ..., p_L)$. Here, $x$ and $y$ are plaintexts, $L$ is the length of plaintext values, and $Enc$ denotes the encryption operation of HE.

\begin{itemize}
\item \textbf{HE-Add:} $ct_x + ct_y = \text{Enc}(x_0 + y_0, x_1 + y_1, ..., x_L + y_L)$
\item \textbf{HE-Mult:} $ct_x \times ct_y = \text{Enc}(x_0 \cdot y_0, x_1 \cdot y_1, ..., x_L \cdot y_L)$
\item \textbf{HE-CMult:} $ct_x \times pt = \text{Enc}(x_0 \cdot p_0, x_1 \cdot p_1, ..., x_L \cdot p_L)$
\item \textbf{HE-Rot:} $\text{Rot}(ct_x, i) = \text{Enc}(x_i, x_{i+1}, ..., x_L, x_0, ..., x_{i-1})$
\end{itemize}

HE multiplication and rotation dominate runtime. Table~\ref{tab:hecost} shows that HE operations are orders of magnitude slower than plaintext, making ciphertext multiplication and rotation the key performance bottlenecks. 

Due to this overhead, efficient HE algorithms must reduce the number of HE multiplications and rotations. This motivates compact data layouts, such as our sparse matrix encoding, to minimize these expensive operations.

\begin{table}[htbp] 
\centering 
\small
\caption{Comparison of computational cost for HE vs. plaintext operations. ``CC" and ``CP" are the multiplications between two ciphertexts, and between ciphertext and plaintext, respectively. “Ratio’’ quantifies the relative cost of HE versus plaintext evaluation.}
\begin{tabular}{c|cc|c} 
\textbf{Operations (ms)} & \textbf{HE} & \textbf{Plaintext} & \textbf{Ratio}\\ 
\hline 
\textbf{Encryption} & 5.501 & - & - \\ \textbf{Decryption} & 2.570 & - & - \\ \textbf{Addition} & 0.550 & 0.009 & 61.1 \\ \multirow{2}[0]{*}{\boldmath{}\textbf{Multiplication}\unboldmath{}} & 20.874(CC) & 0.035 & 596.4 \\ & 4.138(CP) & 0.035 & 118.2 \\ \textbf{Rotation} & 5.350 & 0.130 & 41.15 \\ 
\end{tabular}%
\label{tab:hecost}%
\end{table}%

\subsection{HE-based Dense Matrix-Vector Multiplication}

Matrix-vector multiplication using HE has been extensively studied, primarily focusing on dense matrices for applications in privacy-preserving machine learning and secure data analytics. Although these approaches have improved the efficiency of encrypted computation for dense workloads, they are fundamentally unsuitable for sparse matrix-vector multiplication (SpMV) due to redundant operations and lack of sparsity exploitation. These works fall into two categories: plaintext-ciphertext and ciphertext-ciphertext approaches.

\subsubsection{Plaintext-Ciphertext Matrix-Vector Multiplication} 

GAZELLE~\citep{juvekar2018gazelle}, Delphi~\citep{srinivasan2019delphi}, Cheetah~\citep{huang2022cheetah}, and Rhombus~\citep{he2024rhombus} are representative works that leverage plaintext matrix ciphertext vector multiplication to support efficient secure inference. All of these systems assume that the weight matrix remains in plaintext, while the vector is encrypted, which allows them to exploit faster and more flexible operations supported by HE.


GAZELLE~\citep{juvekar2018gazelle} handles dense data using SIMD-encoded ciphertexts and relies on garbled circuits for nonlinear layers, but does not address sparsity. Delphi~\citep{srinivasan2019delphi} improves computational efficiency by optimizing packing strategies for plaintext-ciphertext products, yet it also targets dense settings. Cheetah~\citep{huang2022cheetah} introduces communication-efficient two-party inference and supports sparse matrices by optimizing the plaintext layout, but still assumes the matrix is in plaintext. Rhombus~\citep{he2024rhombus} focuses on sparse matrix-vector multiplication in two-party inference and leverages structured sparsity and data layout optimizations to accelerate plaintext-ciphertext SpMV.

However, none of these approaches fully supports HE-based ciphertext-ciphertext SpMV, which is the focus of our work. Since one operand is in plaintext in their design, they can use rotation-heavy optimizations and batching techniques that are not directly applicable when both operands are encrypted. Therefore, while these systems are efficient in their targeted scenarios, they cannot be directly applied to our setting, where stronger privacy {guarantees} require that both matrix and vector remain encrypted. Our ciphertext-ciphertext design supports this higher privacy requirement but requires new optimizations to overcome the limited operations and overhead inherent to homomorphic encryption.

\subsubsection{Ciphertext-Ciphertext Matrix-Vector Multiplication}

Falcon~\citep{xu2023falcon} introduces a ciphertext-ciphertext matrix-vector multiplication scheme optimized for homomorphically encrypted convolutional layers. It employs a zero-aware greedy packing algorithm and a communication-aware operator tiling strategy to enhance packing density and reduce latency in homomorphic encryption-based two-party computation frameworks. While Falcon demonstrates significant efficiency improvements, its zero-aware optimization is specifically designed for convolutional layers and not generalized to arbitrary sparse matrices. The main design goal of our proposed approach is to support arbitrary sparse matrices.

Iron~\citep{hao2022iron} introduces an alternative ciphertext-ciphertext matrix-matrix multiplication approach utilizing polynomial evaluation. However, this approach {faces} challenges when being applied directly to sparse matrix-vector multiplication (SpMV). Polynomial-based schemes typically encode matrices as dense polynomials, enabling homomorphic computations via polynomial evaluation and interpolation. Although well-suited for dense matrices, these schemes can become inefficient with sparse data, as representing sparse matrices in a dense polynomial form may result in unused slots, redundant computations involving zero elements, and decreased ciphertext packing efficiency.

HETAL~\citep{lee2023hetal} introduces another way of ciphertext-ciphertext matrix-matrix multiplication, which focuses on privacy-preserving transfer learning with homomorphic encryption by performing encrypted matrix operations for collaborative model training. However, its optimizations are specific to transfer learning contexts and do not support SpMV, limiting their applicability to general sparse linear algebra problems.

HEGMM~\citep{gao2024secure} presents an efficient ciphertext-ciphertext general matrix–matrix multiplication (GEMM) framework under homomorphic encryption, focusing on reducing rotations and ciphertext-ciphertext multiplications through optimized data layouts. While highly effective for dense linear algebra, HEGMM is designed for dense matrix–matrix workloads and does not directly support SpMV. Applying it to sparse matrices would result in massive redundant zero computations and poor packing efficiency.

\subsection{Sparse Matrix-Vector Multiplication (SpMV)} 

Over the past decades, extensive research has focused on optimizing SpMV. Techniques such as compressed storage formats (\eg, CSR, CSC, COO), compute–memory overlap, and hardware specialization have been developed to exploit modern architectures. Recent works have pushed SpMV performance close to hardware limits. For instance, Spaden~\citep{spaden2024} uses bitmap compression and tensor cores to accelerate GPU-based SpMV with 4-bit quantized non-zeros, achieving up to $3\times$ speedup over standard CSR. FastLoad~\citep{fastload2024} decouples memory access via row reordering, and SparseP~\citep{giannoula2022sparsep} performs in-memory SpMV on HBM-PIM platforms for energy-efficient acceleration.

While these advancements are impressive, they are fundamentally tailored to plaintext computation. Naively applying HE techniques to these optimized SpMV kernels introduces critical challenges that break their performance guarantees.

\begin{enumerate}
\item \textbf{Ciphertext size inflation.}
High-performance SpMV kernels rely on compact data layouts and cache-friendly memory access, which are severely disrupted by HE ciphertext inflation. This leads to frequent cache misses, elevated memory pressure, and a sharp decline in throughput.

\item \textbf{Loss of arithmetic efficiency.}
Plaintext SpMV leverages SIMD vectorization and hardware to maximize FLOP/byte ratios. These optimizations are inapplicable under HE, where operations become polynomial multiplications and additions{~\citep{smart2014fully}}. Moreover, HE-specific tasks, such as ciphertext rotation and modulus switching, are computation-intensive and sequential in nature, preventing fusion and parallelism at the hardware level{~\citep{halevi2014algorithms}}. This shifts the performance bottleneck from arithmetic units to cryptographic overhead{~\citep{reagen2021cheetah}}.

\item \textbf{Sparse access patterns require ciphertext rotations.}
Unlike plaintext vectors that support direct indexing, encrypted vectors lack random access. To access specific slots during sparse matrix operations, costly ciphertext rotations are needed. Each row of the sparse matrix may incur an $\mathcal{O}(\log n)$ rotation overhead, which further degrades scalability~\citep{he2024rhombus}.
\end{enumerate}

These limitations underscore that porting plaintext-optimized SpMV kernels to HE settings is impractical. Rethinking sparsity under encryption is crucial for unlocking practical performance in encrypted computation.

\subsection{Difficulty of Implementing SpMV on Non-HE Secure Computing Alternatives}
\label{subsec:nonhe}

\textbf{Secure multi{-}party computation (MPC)}, such as secret sharing or garbled circuits, can evaluate linear algebra exactly without revealing plaintexts to any single party. However, MPC protocols typically require substantial online interaction and bandwidth for each multiplication or AND gate~\citep{MohasselZhang2017SecureML}. This becomes a bottleneck for large, irregular sparse workloads whose communication is driven by the number and placement of non-zeros. In the outsourcing setting targeted here, where both the matrix and the vector are uploaded once under encryption and a semi{-}trusted cloud performs the computation, the required online rounds and the coordination among multiple parties introduce latency and bandwidth sensitivity that are hard to amortize. Our design targets noninteractive evaluation by a single server after the initial upload, which makes MPC a less direct fit for our threat model and deployment assumptions.

\textbf{Trusted Execution Environments (TEEs)}, such as enclave{-}based execution, keep data in plaintext inside hardware{-}protected memory and can offer low arithmetic overhead. Yet they shift trust to hardware and microcode, inherit side{-}channel and rollback concerns, and typically do not obfuscate memory access patterns~\citep{XuCuiPeinado2015ControlledChannel}. Moreover, TEEs do not natively support the requirement that both operands remain encrypted, which is central to our threat model.

\textbf{Hybrid HE with MPC or TEE} often assigns linear parts to homomorphic encryption and handles nonlinearity or control with secure multi{-}party computation or enclaves. These approaches can reduce arithmetic cost and multiplicative depth, but they reintroduce online interaction in the MPC stage or rely on hardware trust in the TEE stage~\citep{juvekar2018gazelle}. In contrast, we target a pure homomorphic encryption pipeline for ciphertext-ciphertext SpMV. After uploading the encrypted  matrix and vector, {the} cloud computing platform evaluates noninteractively and preserves end-to-end confidentiality throughout computation. Our design reduces coordination while addressing the dominant costs in HE, namely rotations and additions, by using the CSSC sparse format for slot alignment and a binary tree aggregation routine, called \textit{totalSum} in our implementation. See Section~\ref{sec:he} for the HE cost model and why minimizing rotations and additions is critical in practice~\citep{halevi2014algorithms}.

\section{Problem Definition and Assumptions}
This section defines the SpMV problem under homomorphic encryption, along with the system architecture and security assumptions. We specify the participating parties, threat model, and accepted structural leakage that enable efficient encrypted computation while keeping all numerical values private.

\subsection{Target: Sparse Matrix-Vector Multiplication}
We focus on the problem of \textit{sparse matrix-vector multiplication (SpMV) under HE}, which is fundamental to many privacy-preserving machine learning and scientific computing applications~\citep{reagen2021cheetah}. We assume that the sparse matrix contains a large number of zero entries, and only the non-zero elements contribute to the result. The positions of both non-zero and zero elements are \textit{publicly known} and shared among all parties involved~\citep{he2024rhombus}, while the values themselves remain private and confidential to other parties.  This setting reflects many real-world use cases where the sparsity pattern is not sensitive.

SpMV typically assumes the input vector is dense, because any sparse vector can be converted to a dense one by removing its zero entries along with the corresponding columns of the matrix.

\subsection{System Model}
We consider a three-party computation model:
\begin{itemize}
  \item \textbf{Client A} owns the sparse matrix and converts it into a compressed format. After compression, the matrix is encrypted using HE and then transmitted to the cloud for further processing.
  \item \textbf{Client B} owns the dense vector and reorganizes it based on the column index information received from Client A. The reorganized vector is then encrypted using HE and transmitted to the server for secure computation.
  \item \textbf{A server}, hosted in the cloud or by a third party, receives the encrypted matrix and vector, performs all HE operations (multiplication, rotation, and constant multiplication), and returns the encrypted result.
\end{itemize}
The server can be independent of A and B, or co-located with either A or B, as long as data privacy is maintained via encryption. This setting enables offloading expensive homomorphic computation to a semi-trusted infrastructure while preserving end-to-end confidentiality~\citep{reagen2021cheetah}.

\subsection{Security Model and Leakage Analysis}
In the following, we delineate which structural information must be revealed to enable correct and efficient SpMV evaluation under homomorphic encryption, and which data remain fully hidden. 

Our HE-based SpMV protocol assumes the standard semi-honest (honest-but-curious) model: the cloud server follows the protocol but may attempt to infer information from received ciphertexts. Clients A and B also follow the protocol, do not collude with the cloud, and do not collude with each other. Under these assumptions, while all values sent to the cloud are encrypted under HE, the cloud {does not learn the locations of non-zero entries, row or column indices, the row-reordering map, intermediate partial sums, or any plaintext information. Structural metadata required for vector reorganization is exchanged only between Client A and Client B.}

Client A sends the CSSC ColumnIndex array in plaintext to Client B; consequently, Client B may learn the matrix’s column-wise sparsity pattern. This leakage profile matches many practical settings in which structural topology is public or less sensitive, while numerical values are sensitive. For example, in road or power-grid networks with public connectivity but private edge weights, their stencil structures are published, and graph degree distributions are known while edge weights or labels remain private. In such settings, exposing sparsity enables substantial performance gain over fully oblivious designs. 

We note, however, that in some other cases, such as patient-specific EHR incidence matrices, or proprietary dependency graphs, the sparsity pattern itself (\ie, which entries are non-zero) may be the primary confidential object.

\subsection{Homomorphic Encryption Strategy}
HE operations are executed exclusively using a SIMD-style packing strategy~\citep{smart2014fully}. Among the two mainstream directions in HE-based computation: \textit{polynomial multiplication} and \textit{element-wise multiplication} (often implemented using SIMD-style packing), we adopt the latter due to its superior efficiency for sparse matrix operations. Polynomial-based methods such as those used in~\citep{reagen2021cheetah} often lead to under-utilization of ciphertext slots and result in extra computations that inflate runtime and memory with sparse matrix manner. In contrast, SIMD packing~\citep{smart2014fully} enables element-wise multiplication by encoding multiple plaintext values into a single ciphertext, although it requires additional operations such as HE-Rotation and HE-CMult. Note that such a SIMD technique for HE differs in meaning from classical hardware SIMD. Our method leverages this packing capability by aligning non-zero matrix entries with their corresponding vector elements, thereby avoiding redundant computations on zero elements and reducing the need for costly HE-Rotation and HE-CMult. Importantly, for each aligned pair of ciphertexts from the matrix and the vector, our method performs only a single HE-Mult operation, ensuring minimal multiplicative depth and significantly improving efficiency.

\section{Compressed Sparse Sorted Column and Matrix-Vector Multiplication Algorithm}

To address the limitations discussed in the previous section, we propose a new sparse matrix format called \textit{Compressed Sparse Sorted Column (CSSC)}. This section first reviews conventional formats such as CSR, CSC, and COO, and then introduces the CSSC structure, followed by a complete matrix-vector multiplication algorithm tailored for encrypted computation.

\subsection{Existing Compressed Formats for Sparse Matrices} 
\label{sec:existCSR}
\begin{figure}
    \centering
    \includegraphics[width=0.95\linewidth]{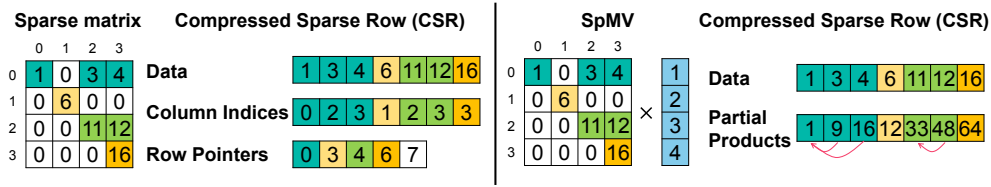}
    \caption{Demonstration of SpMV in the CSR format. When both matrix and vector are encrypted in HE, although partial products can be obtained by aligning elements from the CSR matrix and the vector, aggregating these products is highly inefficient. Each addition requires one HE rotation and one HE constant multiplication, leading to significant computational overhead.}
    \label{fig:csr}
\end{figure}

Efficient {storage} and processing of sparse matrices are crucial in computing. The Compressed Sparse Row (CSR), Compressed Sparse Column (CSC), and Coordinate (COO) are widely used for sparse matrix representation. As shown in Figure \ref{fig:csr},  CSR stores a sparse matrix using three arrays: data (non-zero values), their column indices, and row start pointers, {and} is more suitable for row-based operations. CSC uses a similar structure: non{-}zero values, row indices, and column start pointers, {and} better supports column-based operations. COO stores row indices, column indices, and values for all non-zero elements as separate arrays.  Recent research has proposed hybrid sparse matrix formats that adapt dynamically to matrix characteristics by combining the advantages of different conventional representations (such as CSR and COO) to enhance computational efficiency and memory utilization~\citep{giannoula2022sparsep}. Nevertheless, these hybrid methods still primarily target plaintext computations and remain inadequate for the unique computational and memory access constraints posed by homomorphic encryption.

\subsection{Why Traditional Sparse Matrix Formats Are Incompatible with HE}

Traditional sparse matrix formats such as CSR, CSC, and COO become inefficient under HE, which imposes strict constraints on memory access patterns and data layout. In HE, operations like ciphertext rotation are computationally expensive, and misaligned data significantly increase the cost of element-wise computation.

For example, the CSC format organizes data by columns but stores only row indices and non-zero values. To perform an SpMV where the matrix and vector are encrypted, we need to correctly align elements within ciphertext slots. Since CSC does not explicitly store column indices for each non-zero entry, reconstructing them requires indirect indexing, which complicates slot alignment and increases overhead.

The CSR format, in contrast, stores column indices and non-zero elements in row-major order. This is better for slot-wise alignment during element-wise multiplication, but it introduces additional computational cost during the aggregation step. Specifically, after the element-wise products are computed, the partial results for each row must be summed together. Under HE, this summation requires multiple ciphertext rotations and additions, operations that are much more expensive than plaintext arithmetic. As shown in Figure~\ref{fig:csr}, aggregating the partial products 1, 9, and 16 requires two HE rotations to align 9 and 16 with 1, as well as HE constant multiplications to mask out irrelevant values and retain only the desired sum. This operation is element-specific and cannot be reused for other elements, since the number of non-zero (NNZ) entries per row is unpredictable.

The COO format stores each non-zero entry as a (row, column, value) tuple. Although COO provides complete positional information, it remains inefficient in the HE setting. Similar to CSR, it requires numerous HE operations to aggregate the elements of a row and compute the final result.

\subsection{Definition of the CSSC Format}

\begin{figure}
    \centering
    \includegraphics[width=0.9\linewidth]{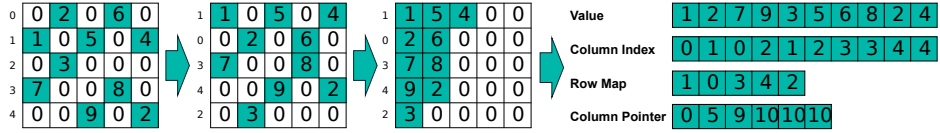}
    \caption{Illustration of the CSSC format: (1) Rows are reordered in descending order according to the number of non-zero elements. (2) Non-zero elements in each row are shifted to the left. (3) Values are then extracted in column-major order, along with their original column indices, row mapping, and column pointers.}
    \label{fig:cssc}
\end{figure}

To efficiently support sparse matrix computation in HE, we propose a new format called Compressed Sparse Sorted Column (CSSC). Figure~\ref{fig:cssc} illustrates the transformation process from a regular sparse matrix into its CSSC format. The matrix is first reordered so that rows appear in descending order of their non-zero counts. Each row’s non-zero elements are then shifted leftward while preserving their original column positions. Finally, the aligned matrix is scanned in column-major order to extract the value array, the corresponding original column indices, the row-mapping permutation, and the column-pointer boundaries. These steps define the structural conventions underpinning the CSSC format.

Formally, let $M \in \mathbb{R}^{r \times c}$ be a sparse matrix after the row-reordering and left-alignment steps described earlier. Its \emph{Compressed Sparse Sorted Column (CSSC)} representation is a tuple
\[
\mathrm{CSSC}(M) = ({VA}, CI, RM, CP),
\]
where:

\begin{itemize}
    \item \textbf{Value Array} ${VA} \in \mathbb{R}^{\mathrm{nnz}(M)}$  
    stores all non-zero entries of the left-aligned triangular matrix in \emph{column-major order}, where $\mathbb{R}$ denotes the set of values (\eg, in real numbers) and $\mathrm{nnz}(M)$ is the number of non-zero elements in $M$.

    \item \textbf{Column Index Array} $CI \in \mathbb{Z}^{\mathrm{nnz}(M)}$ stores the \emph{original column index} of the corresponding non-zero element in $M$, where $\mathbb{Z}$ denotes the set of integers. 
    
    \item \textbf{Row Map Array} $RM \in \mathbb{Z}^{r}$  
    records the original row index in $M$ for each row in the sorted triangularized matrix, where $r$ is the number of rows.
    
    \item \textbf{Column Pointer Array} $CP \in \mathbb{Z}^{c+1}$ is
    a prefix-sum array defined by
    \[
    CP[0] = 0, \qquad CP[j+1] = CP[j] + \mathrm{nnz\_col}(j),
    \]
    where $\mathrm{nnz\_col}(j)$ denotes the number of non-zero elements in column $j$ of the left-aligned triangular matrix, and $c$ is the number of columns.
   
\end{itemize}

By enforcing a structured and column-major layout, CSSC improves ciphertext slot utilization and substantially lowers the number of required rotations and ciphertext-level operations, which constitute the primary performance bottlenecks in homomorphic encryption.

Algorithm~\ref{alg:csrtocssc} in \ref{apx:alg} gives the step-by-step procedure for converting a matrix in CSR representation into CSSC format. Matrices stored in alternative formats could be transformed in a similar way. 

\begin{figure*}
    \centering
    \includegraphics[width=1.0\linewidth]{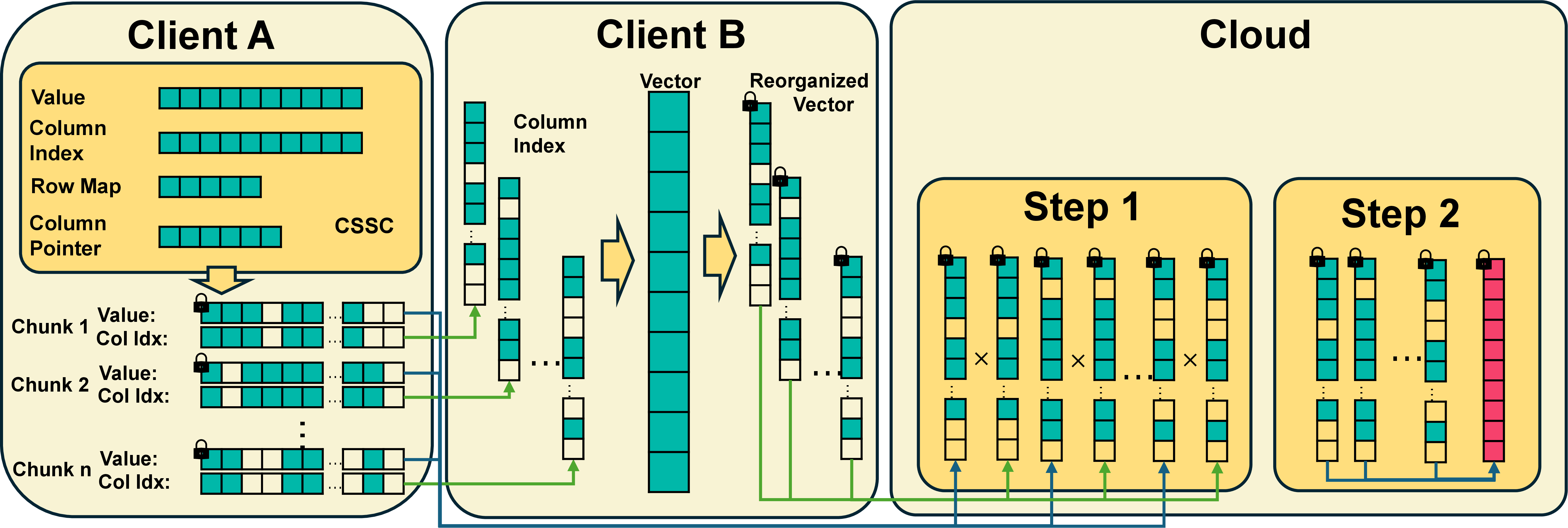}
    \caption{Secure computation framework for sparse matrix–vector multiplication. Client A uploads the encrypted CSSC value array to the cloud. Client B reorganizes its vector using the column index and uploads the encrypted aligned vector. The cloud performs: (1) ciphertext--ciphertext multiplication within each chunk, and (2) aggregation across rows and chunks to produce the final encrypted output.}
    \label{fig:frame}
\end{figure*}

\subsection{Secure Computation Framework of HE-based SpMV}

Our algorithm operates on the proposed \compressformat. As shown in Figure~\ref{fig:frame}, Client A first transforms a sparse matrix into the CSSC format, encrypts the ``Value'' field, uploads it to the cloud, and then shares the unencrypted ``Column Index'' with Client B. Client B reorganizes its vector based on this column index, encrypts it, and then uploads it to the cloud. The cloud server then performs homomorphic multiplication on the corresponding ciphertext pairs from A and B. Finally, {the cloud} aggregate{s} the related elements to get the final result. We summarize the whole process in Algorithm~\ref{alg:spmvalgo}.

\begin{algorithm}[htbp]
\small
\caption{Sparse Matrix-Vector Multiplication}
\label{alg:spmvalgo}
\begin{algorithmic}[1]
\INPUT {CSSC of matrix $M$, Vector $v$}
\OUTPUT {Vector $Res$}

\text{[Client A]}

\text{/*\textit{defined in Algorithm~\ref{alg:chunk_sparse}}*/}

\STATE $value, column\_index,row\_map, param=$ $\text{generateChunk}(M)$ 
\STATE $ct\_value = \text{encrypt}(value)$
\STATE send $ct\_value$ to cloud.
\STATE send $column\_index$ to Client B.

\text{[Client B]}

\text{/*\textit{defined in Algorithm~\ref{alg:getprmedvec}}*/}
\STATE $reorg\_vector=\text{reorgVector}(v, column\_index)$

\STATE $ct\_reorg\_vector = \text{encrypt}(reorg\_vector)$
\STATE send $ct\_reorg\_vector$ to cloud.

\text{[Cloud]}
\STATE Initialize $ct\_ProdList \gets [\,]$
\FOR{$i = 0$ \textbf{up to} $length(ct\_value)$}
    \STATE ${ct\_ProdList[i]} = ct\_value[i] * ct\_reorg\_vector[i]$  \text{/*\textit{HE-Mult}*/}
\ENDFOR

{
\text{/*\textit{defined in Algorithm~\ref{alg:aggregation}}*/}
\STATE $ct\_res = \text{Aggregation}(ct\_ProdList, param.rows, param.cols)$}
    
\STATE send $ct\_res$ to Client A.

{\text{[Client A or Client B (Secret-Key Holder)]}}
\STATE $mid\_res=\text{decrypt}(ct\_res)$
{\STATE $idx = 0$}
\FOR{$rm\_idx$ \textbf{in} $row\_map$}
    \STATE $Res[rm\_idx]=mid\_res[idx]$
    {\STATE $idx=idx+1$}
\ENDFOR

\RETURN $Res$

\end{algorithmic}
\end{algorithm}

\subsubsection{Client A: Chunk Generation}

\begin{figure}
    \centering
    \includegraphics[width=0.93\linewidth]{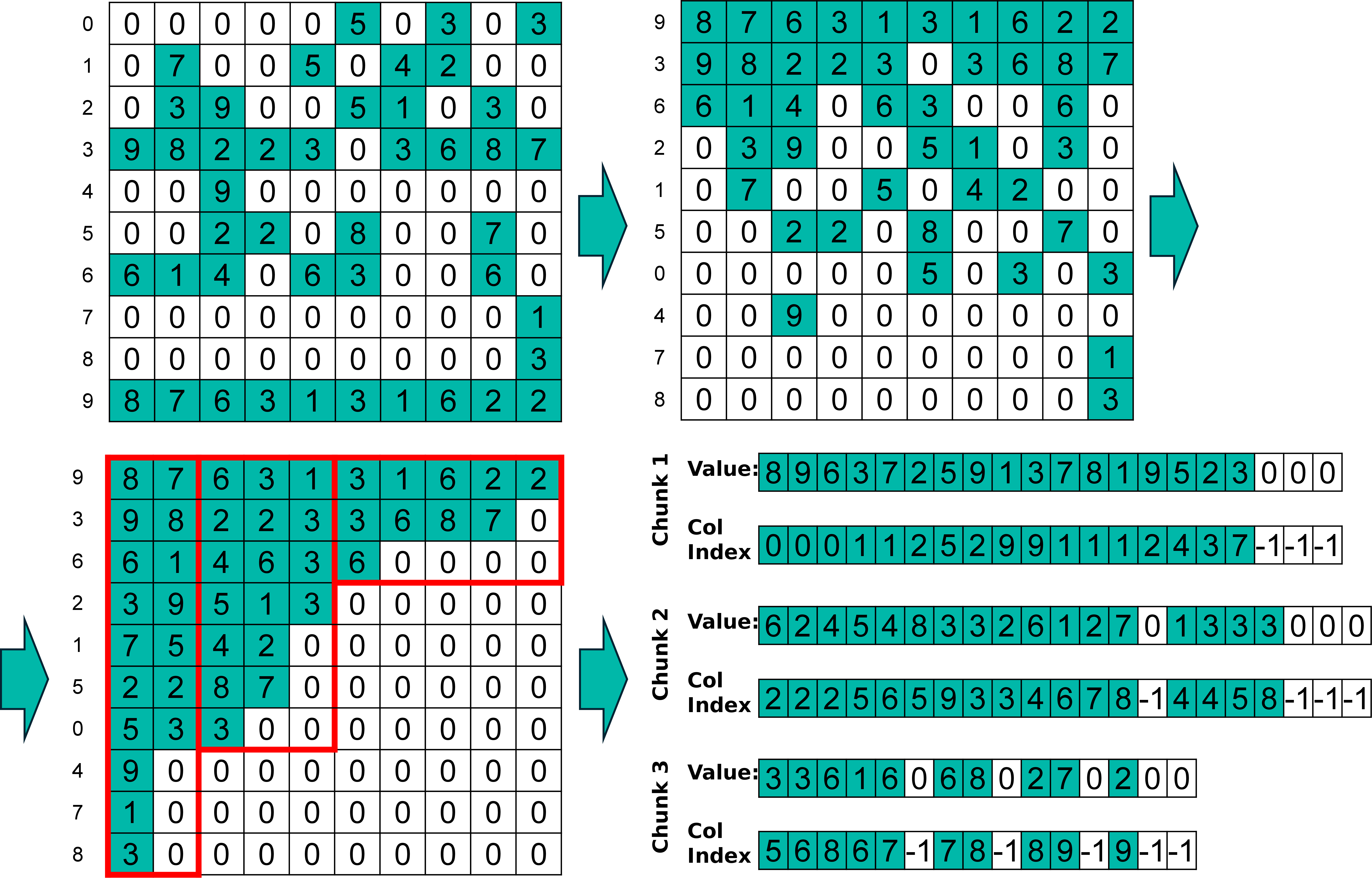}
    \caption{
The process of generating CSSC's value and column index and its chunks. 
(1) Rows are reordered in descending order based on the number of non-zero elements. 
(2) Non-zero elements in each row are shifted to the left. 
(3) In the chunking step, the matrix is divided into chunks. In this example, the three red boxes indicate three chunks, where non-zero and zero elements in the chunks will be preserved, whereas zero elements outside the red boxes (chunks) will be ignored.
}
    \label{fig:chunk21}
\end{figure}

Client A splits the sparse matrix in \compressformat\ into smaller chunks, each of which fits within the size limit of a single ciphertext. To better illustrate the algorithm, we use an upper triangular matrix instead of the \compressformat\ structure to demonstrate the chunking process. Due to that the case in Figure~\ref{fig:cssc} is too small to show the chunking method, we choose a bigger matrix as shown in Figure~\ref{fig:chunk21}. 

In the example of Figure~\ref{fig:chunk21}, we assume a ciphertext size is 21. Consequently, the upper triangular matrix in this example is split into three chunks, ensuring that none exceeds the size limit of 21. The principle behind this process is to fill the ciphertext in a column-major order from left to right until the size limit is reached. Each chunk has a rectangular shape, with its length determined by the maximum number of non-zero elements in all columns within the chunk. 
Within each chunk, the matrix values are arranged in column-major order (flattened), and the corresponding column indices are flattened in the same way. At the same time, the corresponding column index is saved in the {CSSC} format.
This process continues until all columns are processed. The result is a set of compact chunks that are efficient to work with for encrypted computation or storage.

Next, we employ the HE library to encrypt the values and pass the column index to cloud for multiplication. This will be discussed below.

Algorithm~\ref{alg:chunk_sparse} shows the CSSC chunk generation below.

\begin{algorithm}[htbp]
\caption{generateChunk}
\label{alg:chunk_sparse}
\begin{algorithmic}[1]
\INPUT CSSC format: Value list $V$, Column Index list $CI$, Column Pointer list $CP$, Chunk size $s$
\OUTPUT List of chunks: each as ($Value$, $ColumnIndex$), list of row counts, list of column counts

\STATE Initialize $chunks \gets [\,]$, $row\_counts \gets [\,]$, $col\_counts \gets [\,]$
\STATE Initialize $i \gets 0$, $n_C \gets$ number of columns
\STATE Compute the number of non-zeros per column: \\
\hspace{1em} {\textbf{for} $j = 0$ to $n_C-1$: $NNZ[j] \gets CP[j+1] - CP[j]$}

\WHILE{$i < n_C$}
    \STATE $start \gets i$, $count \gets 0$
    
    \text{/*\textit{determine the range of one chunk}*/} 
    \WHILE{$count + NNZ[i] \leq s$}
        \STATE $count \gets count + NNZ[i]$
        \STATE $i \gets i + 1$
    \ENDWHILE

    \STATE $Value \gets$ values(V) from columns $start$ to $i-1$
    \STATE $ColumnIndex \gets$ Column Index(CI) from columns $start$ to $i-1$
    
    \text{/*\textit{$h$ repensents the number of rows in this chunk}*/} 
    \text{/*\textit{by getting the max number of non-zero values}*/}
    \STATE $h \gets NNZ[start]$ 
    \STATE Pad each column with $0$ ($Value$) and $-1$ ($ColumnIndex$) to height $h$
    \STATE Concatenate the padded values into $chunk$
    \STATE Append $chunk$ to $chunks$
    \STATE Append $h$ to $row\_counts$, and $(i - start)$ to $col\_counts$
\ENDWHILE

\RETURN $chunks$, $row\_counts$, $col\_counts$
\end{algorithmic}
\end{algorithm}

\subsubsection{Client B: Reorganization of Dense Vector}
\begin{figure}
    \centering
    \includegraphics[width=0.4\linewidth]{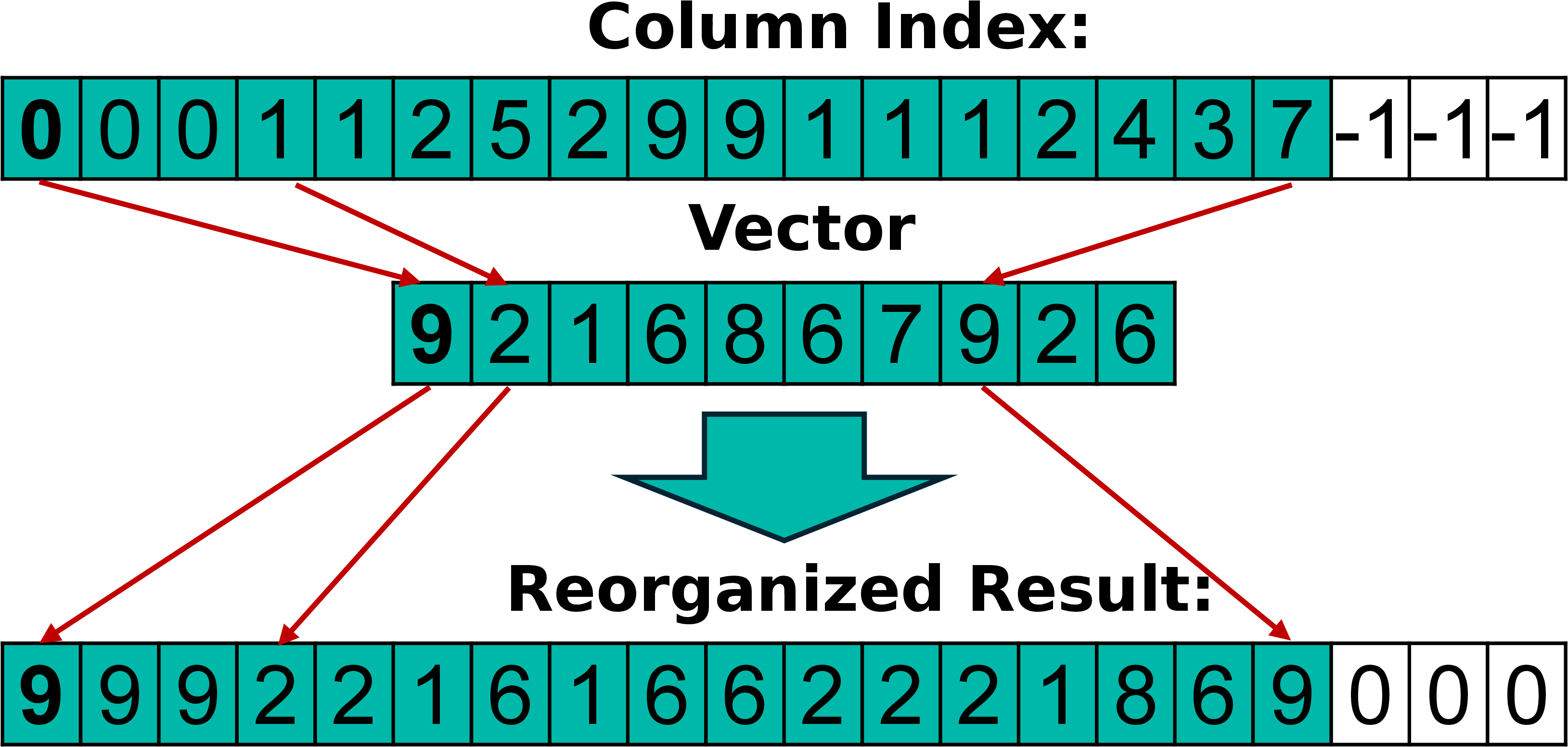}
    \caption{Client B's vector is reorganized using the column indices provided by Client A.
    Each item in ``Column Index" indicates which element in ``Vector" will be placed in the corresponding location of reorganized result, \ie, ReorganizedResult[$k$] = Vector[ColumnIndex[$k$]]. For example, ColumnIndex[3]==1  indicates ``2" (\ie, Vector[1]) will be placed in ReorganizedResult[3].}
    \label{fig:vector}
\end{figure}

To conduct matrix vector multiplication, Client B reorganizes the dense vector according to the column index array generated from the \compressformat\ format by Client A, as shown in Figure~\ref{fig:vector}. We summarize the algorithm in Client B in Algorithm~\ref{alg:getprmedvec}. 

\begin{algorithm}[htbp]
\caption{reorgVector}
\label{alg:getprmedvec}
\begin{algorithmic}[1]
\INPUT 1-D vector $v$, Index list $multip\_col\_idx$, 
\OUTPUT List of permuted vectors $reorg\_vector$

\STATE Initialize $reorg\_vector \gets [\,]$

\FOR{each $col\_idx$ in $multip\_col\_idx$}
    \STATE Initialize $one\_vec \gets [\,]$
    \FOR{each $idx$ in $col\_idx$}
        \IF{$idx \geq 0$}
            \STATE Append $v[idx]$ to $one\_vec$
        \ELSE
            \STATE Append $0$ to $one\_vec$
        \ENDIF
    \ENDFOR
    \STATE Append $one\_vec$ to $reorg\_vector$
\ENDFOR

\RETURN $reorg\_vector$

\end{algorithmic}
\end{algorithm}

\subsection{Cloud Server: Multiplication and Aggregation}

After the cloud server receives the matrix ciphertexts in chunked form along with the reordered vector, it performs homomorphic ciphertext–ciphertext multiplications between each matrix chunk and the vector. Thus, we obtain a list of result chunks, notably, each of which is still in ciphertext. 

To obtain the final result, we need to aggregate these result chunks, which includes two steps: 1) add the corresponding elements together {\it within} each result chunk; 2) add the corresponding elements together {\it across} result chunks.

\begin{figure}[t]
    \centering
    \includegraphics[width=0.83\linewidth]{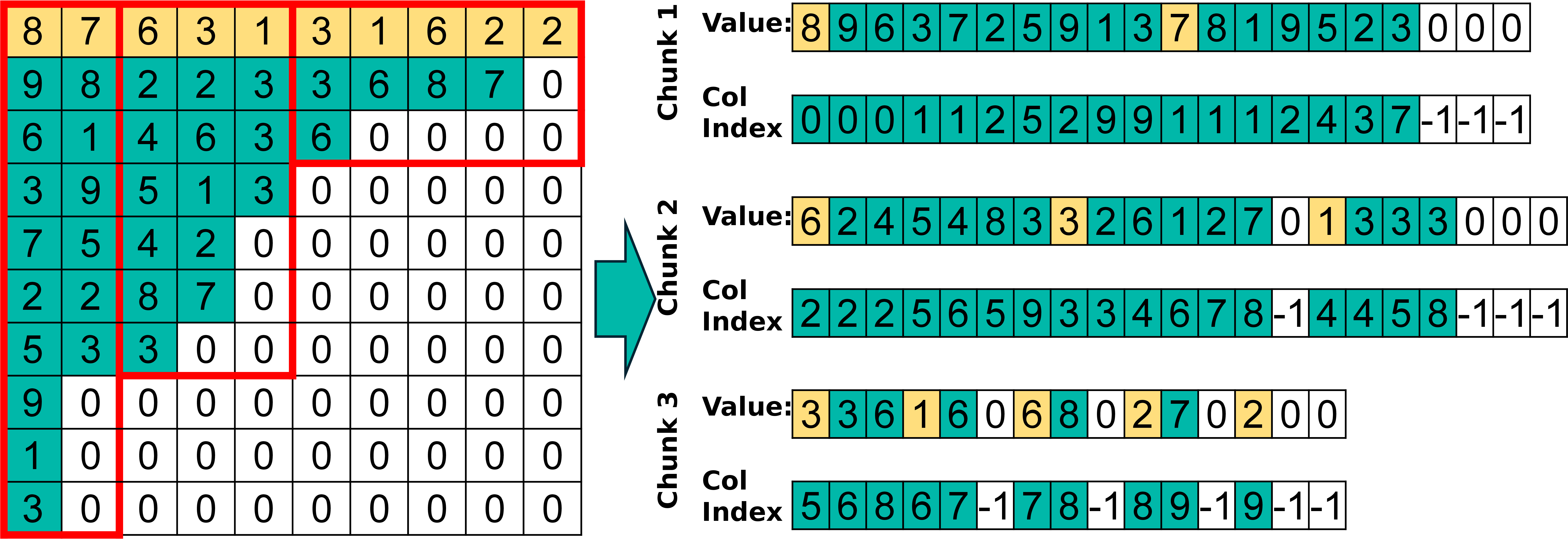}
    \caption{We add all the yellow elements together to get the first elements in result. For chunk 1, it needs to rotate \textbf{10} slots so that 8 can align with 7. For chunk 2, it needs to rotate \textbf{7} slots while it needs \textbf{3} for chunk 3. The values \textbf{10}, \textbf{7}, and \textbf{3}, which represent the number of rows in each chunk, are stored in a list. The corresponding variable in Algorithm~\ref{alg:aggregation} is denoted as \textbf{rList}.}
    \label{fig:addincipher}
\end{figure}

\begin{figure}[htbp]
    \centering
    \includegraphics[width=0.7\linewidth]{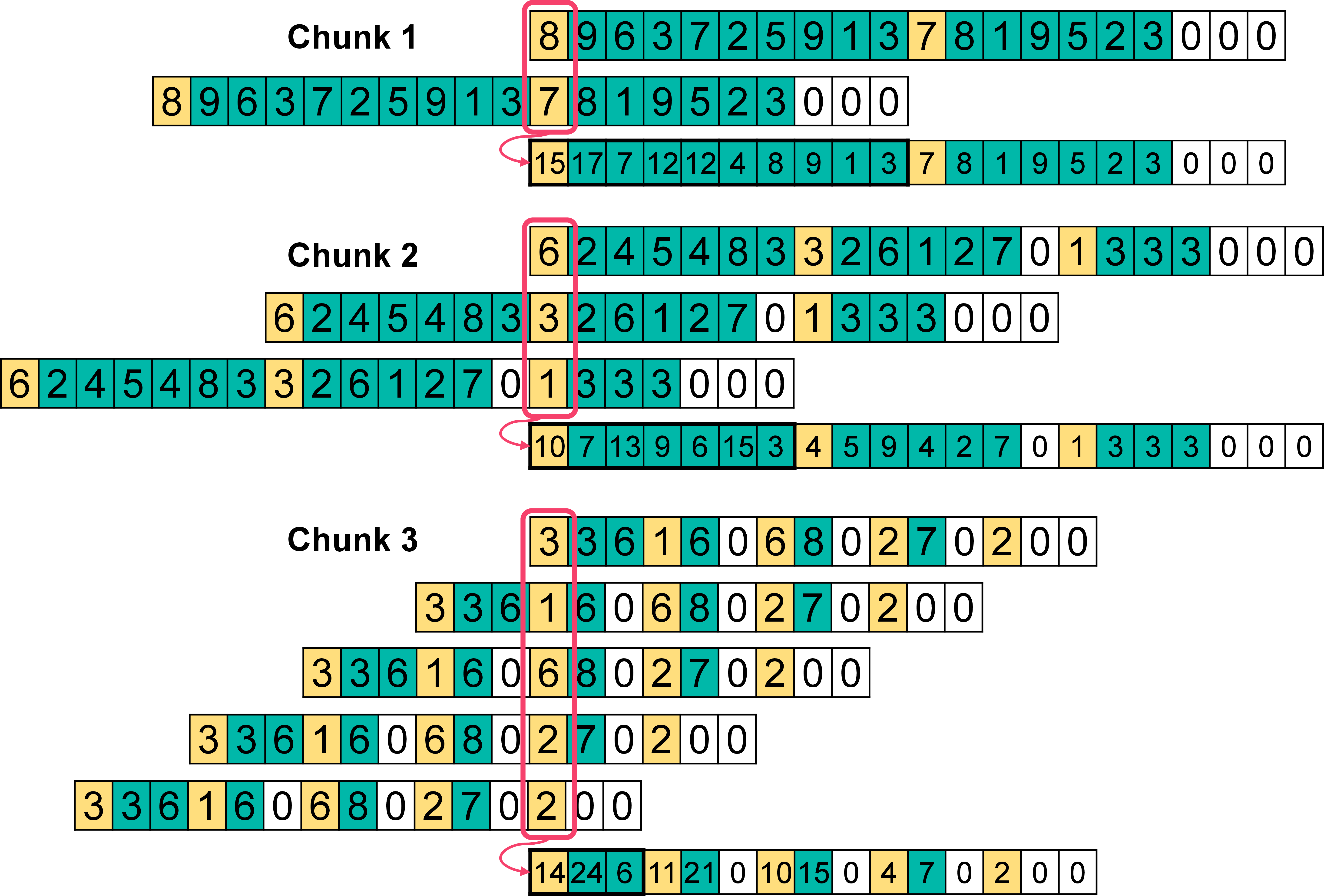}
    \caption{Aggregation Process on Cloud Server. Each chunk is a flattened vector of partial products, which are the result of step 1 in Figure 4. Homomorphic rotations and additions are applied within each chunk to align and accumulate partial sums. Chunks are processed independently prior to global aggregation. Chunk 1 rotates \textbf{1} time, chunk 2 rotates \textbf{2} times and chunk 3 rotates \textbf{4} times. The number \textbf{1}, \textbf{2} and \textbf{4} are the number of column of each chunk, which are denoted as \textbf{cList} in Algorithm~\ref{alg:aggregation}.}
    \label{fig:aggregation-step1}
\end{figure}

As shown in Figure~\ref{fig:addincipher} and~\ref{fig:aggregation-step1}, for chunk 1, it needs to rotate \textbf{10} slots so that 8 can align with 7 and then add the original ciphertext and rotated ciphertext together. For chunk 2, it needs to rotate \textbf{7} slots for each addition between original and rotated ciphertexts, and we need to rotate two times to add all the partial products. For chunk 3, it needs to rotate \textbf{3} slots to align the partial products and needs to rotate four times to get all rotated ciphertexts. Here \textbf{10, 7 and 3} are the number of rows for each chunk. In conclusion, each chunk needs to rotate {\it number of rows} slots and needs to add {\it number of column} elements together. In the above case, the rest of the results in one ciphertext are also added together at the same time.

\begin{figure}[htbp]
    \centering
    \includegraphics[width=0.6
    \linewidth]{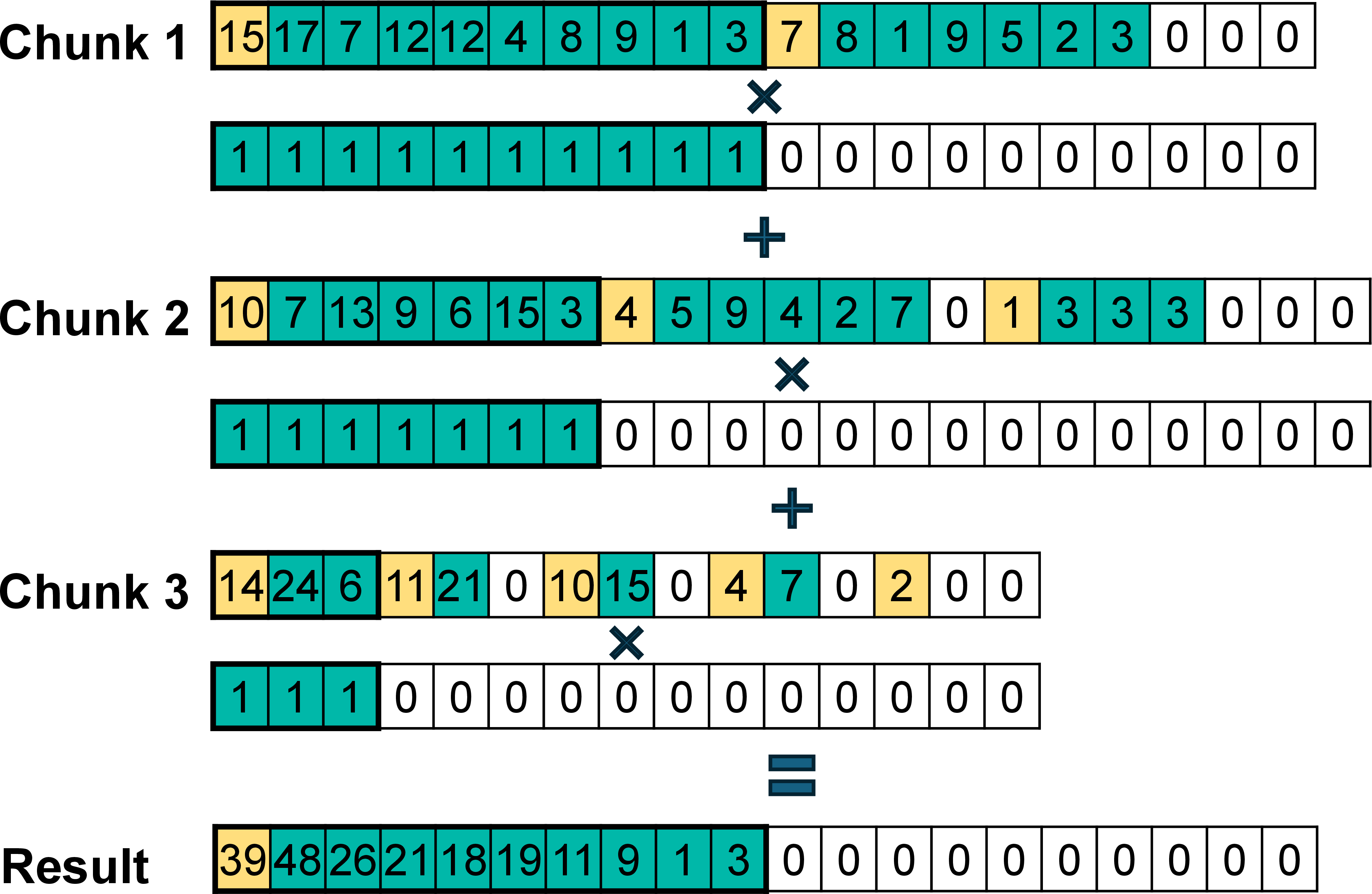}
        \caption{Masking during inter-chunk aggregation. After intra-chunk accumulation, binary masks are applied to ciphertexts to eliminate padded and irrelevant slots. The masked ciphertexts are summed to produce the final encrypted result vector, ensuring correctness when chunks have differing row sizes.}
    \label{fig:aggregation-step2}
\end{figure}

Then, we need to add three chunks together. As shown in Figure~\ref{fig:aggregation-step2}, for Chunk 1, only the first 10 elements are meaningful. For chunks 2 and 3, only the first 7 and 3 elements are useful, respectively. The remaining elements are meaningless. Therefore, we need a mask for each chunk and the mask {is}:
\begin{equation}
    mask = [\underbrace{{1,1,1,...,1}}_{\text{number of rows in this chunk}},0,0,...,0]
\end{equation}
Then, result will be:
\begin{equation}
    result = \sum_i Chunk_i \times mask_i
\end{equation}

We summarize the aggregation procedure in Algorithm~\ref{alg:aggregation}.
The inputs are a list of ciphertexts $\mathbf{ctList}=\{ctV_i\}$ and the
corresponding chunk shapes $(r_i,c_i)$ given by $\mathbf{rList}$ and $\mathbf{cList}$.
Each $ctV_i$ packs a $r_i \times c_i$ intermediate chunk.

\begin{algorithm}[htbp]
\caption{Aggregation}
\label{alg:aggregation}
\small
\begin{algorithmic}[1]
\INPUT Encrypted vector list $\textbf{ctList}$, row sizes $\textbf{rList}$, column sizes $\textbf{cList}$
\OUTPUT {Encrypted result ciphertext $\textbf{res}$}
\STATE Initialize $intraRes \gets [\ ]$
\FOR{each $(ctV, r, c)$ in $(\textbf{ctList}, \textbf{rList}, \textbf{cList})$} 
    \STATE $w \gets ctV$ \text{/*\textit{the ciphertext used for accumulation within a chunk}*/}
    \STATE $e \gets 1$ \text{/*\textit{the step size controlling the rotation offset}*/}
    \FOR{$j = \text{numBits}(c) - 2$ \textbf{down to} $0$}

        \STATE $w \gets w + \text{HE.Rot}(w, e \cdot r)$ 
        \text{/*\textit{rotate ciphertext $w$ by $e \cdot r$ slots}*/}
        \STATE $e \gets 2 \cdot e$
        \IF{$\text{bit}_j(c) = 1$}
            \STATE $w \gets w + \text{HE.Rot}(w, 1 \cdot r)$
            \STATE $e \gets e + 1$
        \ENDIF
    \ENDFOR
    \STATE Append $w$ to $intraRes$
\ENDFOR

\STATE $res \gets 0$
\FOR{$i = 0$ \textbf{to} $\text{length}(intraRes)-1$}
   \STATE $mask \gets \text{HE.Encode}([1]^{rList[i]} \Vert [0]^{N_{\text{slots}}-rList[i]})$
    \STATE $res \gets res + (intraRes[i] \cdot mask)$    
    
\ENDFOR

\RETURN $\textbf{res}$
\end{algorithmic}
\end{algorithm}

\paragraph{Lines 1--13: intra-chunk aggregation}
Line~1 initializes an empty list \texttt{intraRes} to store one aggregated ciphertext per chunk.
For each tuple $(ctV,r,c)$ (line~2), line~3 sets $w\gets ctV$, where $w$ is the working ciphertext used
to accumulate row-wise sums inside the current chunk.
Line~4 initializes $e\gets 1$, where $e$ controls the rotation offset in units of columns. Lines~5--11 implement the \texttt{totalSum} gadget from~\citep{halevi2014algorithms}, which aggregates
slots within a packed ciphertext in $O(\log (n_{\text{slots}}))$ using a doubling strategy.
Here, $\texttt{numBits}(c)$ denotes the bit-length of integer $c$
(\eg, $\texttt{numBits}(5)=3$, $\texttt{numBits}(21)=5$), and $\texttt{bit}_j(c)\in\{0,1\}$
denotes the $j$-th least significant bit.
The loop iterates $j$ from $\texttt{numBits}(c)-2$ down to $0$ (line~5), and line~6 updates
\[
w \leftarrow w + \mathrm{HE.Rot}(w, e\cdot r),
\]
where $\mathrm{HE.Rot}(u,k)$ rotates ciphertext $u$ by $k$ slots.
Because each column occupies $r$ contiguous slots, the rotation by $e\cdot r$ aligns the next $e$
columns with the current $e$ columns, enabling row-wise accumulation.
Line~7 doubles the step size ($e\leftarrow 2e$) to reflect that $w$ now aggregates $2e$ columns.
If $\texttt{bit}_j(c)=1$ (line~8), line~9 performs an additional rotation by $r$ slots to include the
extra column block, and line~10 increments $e$ accordingly ($e\leftarrow e+1$).
After the inner loop terminates, each row of $w$ contains the sum across all $c$ columns of the chunk.
Line~13 appends the aggregated ciphertext $w$ to \texttt{intraRes}.

\paragraph{Lines 15--23: inter-chunk aggregation}
Line~15 sets $maxRow\gets rList[0]$ (\ie, the target row length for the final output).
For each aggregated chunk ciphertext \texttt{intraRes[i]} (line~16),
line~17 constructs a plaintext mask vector consisting of $rList[i]$ ones (followed by zeros if padded),
and encodes it via $\mathrm{HE.Encode}(\cdot)$ so it can be used in ciphertext--plaintext multiplication.
Line~18 multiplies the ciphertext by the encoded mask (denoted by $\mathrm{HE.CMult}$) to zero out padded
slots, and adds the masked ciphertext into the accumulator \texttt{res} (denoted by $\mathrm{HE.Add}$).
Finally, line~20 returns \texttt{res}.

\subsection{Time Complexity}
We analyze the computational cost of the proposed scheme by separating the work performed by the clients from the homomorphic operations executed in the cloud. The following subsections present the time complexity of each component.

\subsubsection{Client}
Let $n_C$ be the number of columns, $\mathrm{NNZ}_{\text{total}} = |{VA}|$ be the total number of non-zero entries, $s$ be the chunk size.

We analyze each part of Algorithm~\ref{alg:chunk_sparse} in turn.

\begin{table}[htbp]
\centering
\renewcommand{\arraystretch}{1.2}
\caption{Complexity summary for getting chunk from CSSC (Algorithm~\ref{alg:chunk_sparse}).}
\begin{adjustbox}{max width=\linewidth}
\begin{tabular}{l c c}
\textbf{Operation} & \textbf{Time Complexity} & \textbf{Alg.\ \ref{alg:chunk_sparse} Line(s)} \\
\hline
NNZ counting (per column) 
  & $\mathcal{O}(n_C)$ 
  & 3 \\

Column grouping (chunking) 
  & $\mathcal{O}(n_C)$ 
  & 4--17 \\

Value extraction 
  & $\mathcal{O}(\mathrm{NNZ}_{\text{total}})$ 
  & 10--11 \\

Padding and concatenation 
  & $\mathcal{O}(\mathrm{NNZ}_{\text{total}} + n_C s)$ 
  & 13--14 \\

Metadata updates 
  & $\mathcal{O}(\#\text{chunks})$ (negligible) 
  & 16 \\

\hline
\textbf{Total} 
  & $\mathcal{O}(\mathrm{NNZ}_{\text{total}} + n_C s)$ 
  & --- \\

Simplified 
  & $\mathcal{O}(\mathrm{NNZ}_{\text{total}})$ 
  & --- \\

\end{tabular}
\end{adjustbox}
\end{table}

The NNZ-counting pass is $\mathcal{O}(n_C)$, and the chunking loops advance the
column index monotonically, giving another $\mathcal{O}(n_C)$ overall. Extracting
values touches each non-zero exactly once, so $T_{\text{extract}}
=\mathcal{O}(\mathrm{NNZ}_{\text{total}})$. Padding is bounded by the chunk
height $s$, yielding at most $N_{\text{pad}}\le n_C s -
\mathrm{NNZ}_{\text{total}}$ additional entries and total padding cost
$\mathcal{O}(\mathrm{NNZ}_{\text{total}}+n_C s)$. Metadata updates are
$\mathcal{O}(1)$ per chunk and negligible. Thus the end-to-end complexity is
\[
T_A=\mathcal{O}(n_C s+\mathrm{NNZ}_{\text{total}}),
\]
which simplifies to $\mathcal{O}(\mathrm{NNZ}_{\text{total}})$, since $n_C s$ is {often} bounded by $2\cdot\mathrm{NNZ}{\text{total}}$.

Client~B constructs the permuted vectors by accessing each non-zero index exactly once. The zero filling step introduces only a constant overhead. Therefore the total cost is $T_B=\mathcal{O}(\mathrm{NNZ}_{\text{total}})$ The pseudocode for the procedure in Client~B is provided in Algorithm~\ref{alg:getprmedvec}.

\subsubsection{Cloud}
\label{sec:cloudtimecom}

In the secure matrix-vector multiplication pipeline, the cloud first performs a ciphertext-ciphertext multiplication between each encrypted matrix chunk and the corresponding reorganized encrypted vector:
\texttt{ct\_res = ct\_value[i] * ct\_reorg\_vector[i]} which produces all the results of products among matrix multiplication. 

\begin{table*}[htbp]
  \centering
  \caption{Time complexity analysis for the algorithm. Here, $n_{\text{ct}}$ is the number of ciphertexts, and $C_{\text{max}}$ is the maximum column size. Step 1 corresponds to the Step 1 in Figure~\ref{fig:frame}, and Step 2 corresponds to the Step 2 in Figure~\ref{fig:frame}. "$1M$" and "$1C$" denote 1 depth from ciphertext-ciphertext and ciphertext-plaintext multiplications, respectively.}

    \begin{tabular}{c|ccccc}
          & \textbf{HE-Mult} & \textbf{HE-CMult} & \textbf{HE-Rot} & \textbf{HE-Add} & \textbf{Depth} \\
    \hline
    \textbf{Step 1} & $n_{\text{ct}}$     & 0     & 0     & 0     & $1M$ \\
    \textbf{Step 2} & 0     & $n_{\text{ct}}$     & $n_{\text{ct}}\cdot \log_2 C_{max}$ & $n_{\text{ct}}\cdot \log_2 C_{max}$ & $1C$ \\
    \textbf{Total} & $n_{\text{ct}}$     & $n_{\text{ct}}$     & $n_{\text{ct}}\cdot \log_2 C_{max}$ & $n_{\text{ct}}\cdot \log_2 C_{max}$ & $1M+1C$ \\
    \end{tabular}%
  \label{tab:timecomp}%
\end{table*}%

To aggregate the resulting partial ciphertexts efficiently, we design the \texttt{Aggregation} algorithm, which minimizes multiplicative depth while enabling scalable homomorphic computation. The algorithm operates in two stages: (1) an intra-ciphertext accumulation using a binary-tree-style rotation and addition scheme, which is from HElib~\citep{halevi2014algorithms}, and (2) an inter-ciphertext aggregation step leveraging ciphertext-plaintext multiplications.

Our operation-level analysis shows that \texttt{Aggregation} incurs $O(n_{\text{ct}} \cdot \log C_{\text{max}})$ homomorphic rotations and additions, where $n_{ct}$ is the number of ciphertexts and $C_{\text{max}}$ denotes the maximum column size among them.

The time complexities are summarized in Table~\ref{tab:timecomp}.

\subsubsection{Noise Growth}

Our BFV implementation uses the 128-bit secure parameter set $(N = 8192,\; t = 65537,\; \log_2 Q = 200)$, which provides an initial noise budget of roughly 146 bits and is sufficient for the shallow arithmetic circuit in our sparse SpMV pipeline. Ciphertext addition and ciphertext rotation contribute only small additive noise, whereas ciphertext-ciphertext multiplication produces the dominant noise growth by approximately doubling the existing noise term. Each chunk in our design performs only one ciphertext-ciphertext multiplication, one ciphertext-plaintext multiplication, and a logarithmic number of additions and rotations, which keeps the multiplicative depth fixed at two and independent of matrix size or sparsity. The empirical behavior is consistent with this analysis. As shown in Table~\ref{tab:mult_noise}, the noise budget decays from 146 bits to 114, 81, 48, and 15 bits across four consecutive ciphertext-ciphertext multiplications, demonstrating that our parameter set can safely support up to four such multiplications, significantly more than our circuit requires. Ciphertext-plaintext multiplication decays even more slowly, retaining 19 bits after five repetitions. These results confirm that the remaining noise budgets in all experiments remain well above the decryption threshold, ensuring correctness without bootstrapping and allowing a single parameter choice to be reused across all datasets.

\begin{table}[h]
\centering
\small
\caption{Noise budget decay (bits) under repeated ciphertext{-}ciphertext and ciphertext{-}plaintext multiplications.}
\begin{tabular}{c|ccccccc}

\textbf{Operation} & \textbf{0} & \textbf{1} & \textbf{2} & 
\textbf{3} & \textbf{4} & \textbf{5} & \textbf{6} \\
\midrule
ct$\times$ct & 146 & 114 & 81 & 48 & 15 & 0 & 0 \\
ct$\times$pt & 146 & 121 & 96 & 71 & 45 & 19 & 0 \\

\end{tabular}

\label{tab:mult_noise}
\end{table}

\subsection{Handling Large Matrices}

When processing large sparse matrices that exceed the ciphertext capacity, our proposed method leverages the flexibility of the compressed sparse sorted column (CSSC) format. To handle large matrices, we horizontally partition the matrix; that is, we split the matrix along the row dimension such that the number of rows in each partition does not exceed the ciphertext slot capacity. This ensures that even the longest chunk generated after compression stays within the ciphertext size limit. By doing so, we maintain efficient ciphertext packing, avoid unnecessary zero-padding, preserve the sparsity of the original data, and finally enable scalable processing of large sparse matrices.

In contrast, to ensure fairness, we preprocess the other approaches by partitioning the large matrix into smaller blocks, aiming to form square submatrices whenever possible. Due to the irregular dimensions of large matrices, it is not always feasible to partition them evenly into perfect squares. In such cases, the remaining rightmost columns or bottommost rows that do not form complete squares are padded with zeros to expand them into square submatrices. We then iteratively apply the baseline encryption and computation to each submatrix and aggregate the results.

\section{Experiments}
All experiments were conducted using matrices from SuiteSparse Matrix Collection \citep{davis2011suitsparse}, a widely recognized repository for sparse matrix benchmarking in scientific computing. We choose a variety of sizes and sparsity structures from the collection to comprehensively evaluate the performance of our proposed method.

The experiments were performed on a machine equipped with an Intel Xeon Silver 4114 CPU (2.20\,GHz, 10 cores) and 256\,GB RAM; {Unless otherwise stated,} all runs were executed in \textbf{single-threaded} mode for consistency and fairness. Our implementation uses Pyfhel~\citep{ibarrondo2021pyfhel} with the BFV scheme at a 128-bit security level and a \emph{single} fixed parameter set for all datasets and baselines: ring degree $N{=}8192$ with batching (8192 slots), plaintext modulus $t{=}65537$ ($t\equiv 1 \bmod 16384$), and a coefficient-modulus chain with bit-lengths $[60,\,40,\,40,\,60]$ (total $\log_2 Q{=}200$ bits); the circuit performs one ciphertext--ciphertext multiplication followed by additions/rotations (no bootstrapping), and we generate one relinearization key set plus powers-of-two rotation keys.

To account for potential system fluctuations, each experiment was independently repeated five times, and the reported results are the averages of these five runs. We primarily report:
\begin{itemize}
    \item \textbf{Execution time}, including encryption and homomorphic computation phases.
    \item \textbf{Peak memory usage} {measured via \texttt{tracemalloc} during the execution.}
    \item \textbf{Ciphertext communication overhead} {reported in MB based on measured ciphertext sizes and the number of ciphertexts transmitted.}
\end{itemize}
Ciphertext communication overhead is a critical metric in HE applications, particularly in cloud-based or distributed environments where encrypted data must be exchanged between parties. Since ciphertexts are significantly larger than plaintexts due to encryption overhead, communication can become a major bottleneck. Measuring this overhead helps assess the practicality and scalability of our approach.

Due to space limitations, representative results for key matrices are presented in the main text, while the complete experimental data is included in \ref{apx:exp}.

\subsection{Baseline Appoaches}
\label{sec:expbase}
\begin{figure}
    \centering
    \includegraphics[width=0.4
    \linewidth]{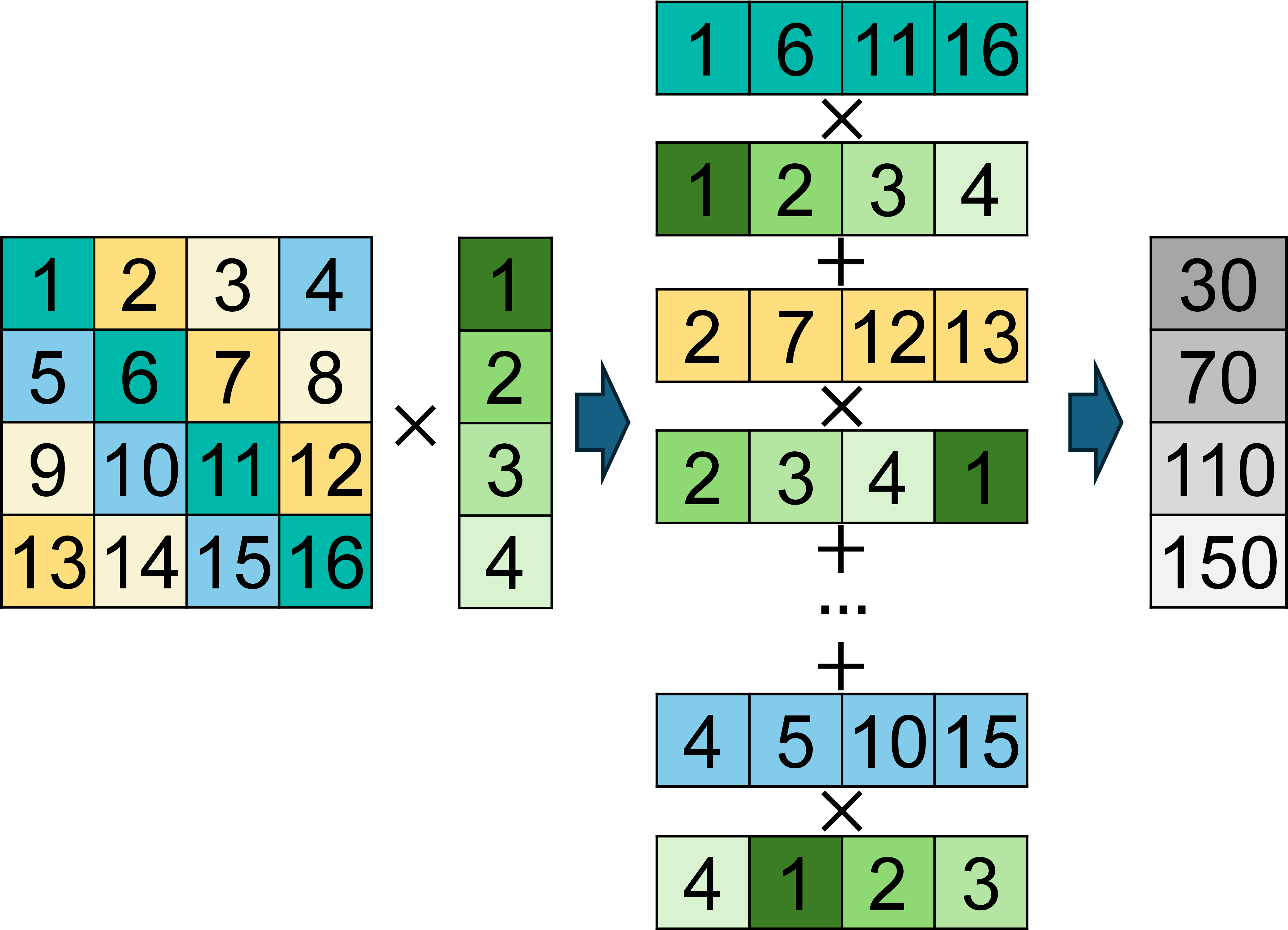}
    \caption{Diagonal method. First extract each diagonal of the matrix and encrypt it into a separate ciphertext. For each encrypted diagonal, perform a homomorphic multiplication with an appropriately rotated ciphertext of the input vector, aligning the vector elements with the corresponding diagonal entries on cloud. Finally, all resulting partial products are homomorphically summed to obtain the final matrix–vector multiplication result.}
    \label{fig:basediag}
\end{figure}

To the best of our knowledge, no prior work directly addresses ciphertext-ciphertext sparse matrix-vector multiplication. Thus, we adopt the straightforward diagonal method~\citep{juvekar2018gazelle} as one baseline approach. As shown in Figure~\ref{fig:basediag}, it encrypts every diagonal as a single ciphertext, uploads it to cloud, multiplies by a rotated vector and aggregates together. The entries in the same color come from the same ciphertext.

We also include HEGMM~\citep{gao2024secure} as a baseline. HEGMM targets ciphertext-ciphertext matrix-matrix multiplication under HE. Its runtime is proportional to the minimum of the matrices dimensions. 
To enable SIMD-style elementwise products via alignment, HEGMM constrains the number of rows or columns to be at most $\sqrt{n}$, where $n$ is the ciphertext slot count, so that an entire row or column fits in one ciphertext. And this is intrinsic to HEGMM. In our experiments, HEGMM does not perform well on sparse matrix, mainly due to the large size of matrices, which require blocking and increases computational complexity. However, HEGMM may offer advantages in small matrix scenarios, particularly when both the number of rows and columns are below $\sqrt{n}$, allowing multiplication to be performed without blocking.

HETAL is not a dedicated SpMV primitive but instead a general homomorphic tiling and packing strategy for ciphertext–ciphertext matrix operations. Although we report HETAL results, in the large, highly sparse setting studied here, it does not provide a direct advantage over the diagonal baseline.

Although Rhombus~\citep{he2024rhombus} works in a different trust model by performing ciphertext to plaintext matrix–vector multiplication, we still include it as an additional reference. Rhombus stores each matrix row as a plaintext vector and multiplies it with the encrypted input vector, then applies a binary tree style rotate and accumulate procedure. This design requires one ciphertext operation for every row and therefore introduces many ciphertext to plaintext multiplications and rotations on tall sparse matrices. It also needs to keep a large number of plaintext rows in memory. Even though this comparison is not in Rhombus’s favor, our method still runs faster and uses less memory because CSSC produces far fewer ciphertexts and reduces the total number of homomorphic operations.

\begin{table}[t]
  \centering
  \caption{Performance comparison between our proposed method and the baseline approaches for sparse matrix operations under homomorphic encryption. Each row reports the matrix dimensions (rows, columns, and number of non-zeros), followed by the execution time (in seconds) for four methods: our proposed method, Diag~\citep{halevi2014algorithms}, HEGMM~\citep{gao2024secure}, and HETAL~\citep{lee2023hetal}. The final column is the speedup factor, computed as the ratio of best execution time among all three baseline methods to that of our method. Entries marked with ``$>3$ days" indicate that the method did not complete within the time limit. The results highlight the significant computational efficiency of our approach.
}
\begin{adjustbox}{max width=\linewidth}
\begin{tabular}{c|ccc|cccc|c}

\multicolumn{1}{c|}{\multirow{2}{*}{\textbf{Matrix}}} &
\multicolumn{1}{c}{\multirow{2}{*}{\textbf{Rows}}} &
\multicolumn{1}{c}{\multirow{2}{*}{\textbf{Cols}}} &
\multicolumn{1}{c|}{\multirow{2}{*}{\textbf{NNZ}}} &
\multicolumn{4}{c|}{\textbf{Time (s)}} &
\multicolumn{1}{c}{\multirow{2}{*}{\textbf{Speedup (x)}}} \\
& & & &
\textbf{Ours} & \textbf{Diag\citep{juvekar2018gazelle}} &
\textbf{HEGMM\citep{gao2024secure}} & \textbf{HETAL\citep{lee2023hetal}} & \\
\midrule
arc130         & 130   & 130    & 1354     & \bfseries 0.25  & 7.52      & 9.83        & 8.33         & 30.08 \\
stat96v5       & 2307 & 75779 & 233921   & \bfseries 17.78 & 5465.94  & 69524.83   & 40436.16    & 307.42 \\
M80PI\_n1      & 4028 & 4028  & 9927     & \bfseries 0.82  & 294.71    & 6317.35    & 3659.79     & 359.40 \\
mycielskian13  & 6143 & 6143  & 1227742 & \bfseries 71.76 & 1030.40  & 14649.79   & 8497.93     & 14.36 \\
nemeth24       & 9506 & 9506  & 1506550 & \bfseries 97.15 & 1859.81  & 35140.14   & 20502.51    & 19.14 \\
msc10848       & 10848& 10848 & 1229776 & \bfseries 81.58 & 56198.48 & 45707.00   & \text{\textgreater 3 days} & 560.28 \\
as-caida       & 26475& 26475 & 53381    & \bfseries 8.77  & 18750.38 & 272714.29  & 159955.17   & 2138.01 \\
ship\_001      & 34920& 34920 & 3896496 & \bfseries 268.77& 21398.58 &  {\textgreater 3 days} & {\textgreater 3 days} & 79.62 \\
cca            & 49152& 49152 & 139264   & \bfseries 14.89 & 17686.71 & {\textgreater 3 days} & {\textgreater 3 days} & 1187.82 \\
p2p-Gnutella31 & 62586& 62586 & 147892   & \bfseries 12.62 & 74553.33 & {\textgreater 3 days} & {\textgreater 3 days} & 5907.55 \\

\end{tabular}
\end{adjustbox}
  \label{tab:time}%
\end{table}%

\subsection{Performance Evaluation}
Table~\ref{tab:time} presents a performance comparison between our proposed method and the baseline approaches for sparse matrix operations under homomorphic encryption. The matrices span a broad range of sizes and sparsity levels, from small cases like \texttt{arc130} (130 × 130 with 1,354 non-zeros) to large-scale matrices such as \texttt{p2p-Gnutella31} (62,586 × 62,586 with 147,892 non-zeros).

Across all tested matrices, our method consistently achieves substantial speedups over the baselines, with improvements ranging from approximately 18× to over 5,000×. 
For instance, on moderately sized matrices such as \texttt{stat96v5} and \texttt{M80PI\_n1}, our method outperforms the fastest baseline by over 300×. On larger matrices, such as \texttt{msc10848} and \texttt{as-caida}, the performance {gains} become even more striking, reaching around 600× and 2,000×, respectively. Notably, for large matrices like \texttt{ship\_001}, \texttt{cca}, and \texttt{p2p-Gnutella31}, the HEGMM and HETAL methods fail to complete within three days, while our method finishes within {the time limit (from sub-second to a few minutes on the tested instances)}, yielding speedups of 80x to over 5,000×.

These dramatic improvements are attributed to two core advantages of our approach: (1) an efficient sparse representation that avoids redundant computation on zero elements, and (2) optimized ciphertext packing and aggregation that significantly reduce the number of homomorphic operations. In contrast, the baseline methods suffer from inefficient handling of sparsity and excessive ciphertext-level overhead.

Overall, these results demonstrate strong scalability and computational efficiency of our method, making it a practical and robust solution for privacy-preserving sparse matrix operations under homomorphic encryption.
More experiment results are shown in {Table~\ref{tab:expapdx} in \ref{apx:exp}.

\subsubsection*{Empirical Validation of Time Complexity for Cloud Server}

To validate the theoretical time complexity of our cloud-side computation, we analyze the relationship between runtime and matrix sparsity. As shown in Section~\ref{sec:cloudtimecom} and Table~\ref{tab:timecomp}, the aggregation step on the cloud incurs a complexity of \( O(n_{\text{ct}} \cdot \log C_{\max}) \), where $n_{ct}$ is the number of ciphertext chunks and \( C_{\max} \) is the maximum number of columns per chunk. Since \( C_{\max} \) remains small and relatively stable across datasets, the runtime is expected to scale approximately linearly with the number of non-zero elements (\( \text{NNZ} \)).

Figure~\ref{fig:loglog_runtime} presents a log-log plot of cloud runtime versus NNZ across a diverse set of evaluated matrices. The near-linear trend in log-log space indicates a power-law relationship with a slope close to 1, confirming that the runtime grows proportionally to NNZ. This strongly supports our theoretical analysis and demonstrates that the proposed framework maintains scalable and efficient cloud-side performance even for large, high-dimensional sparse matrices.

\begin{figure}[h]
  \centering
  \includegraphics[width=0.7\linewidth]{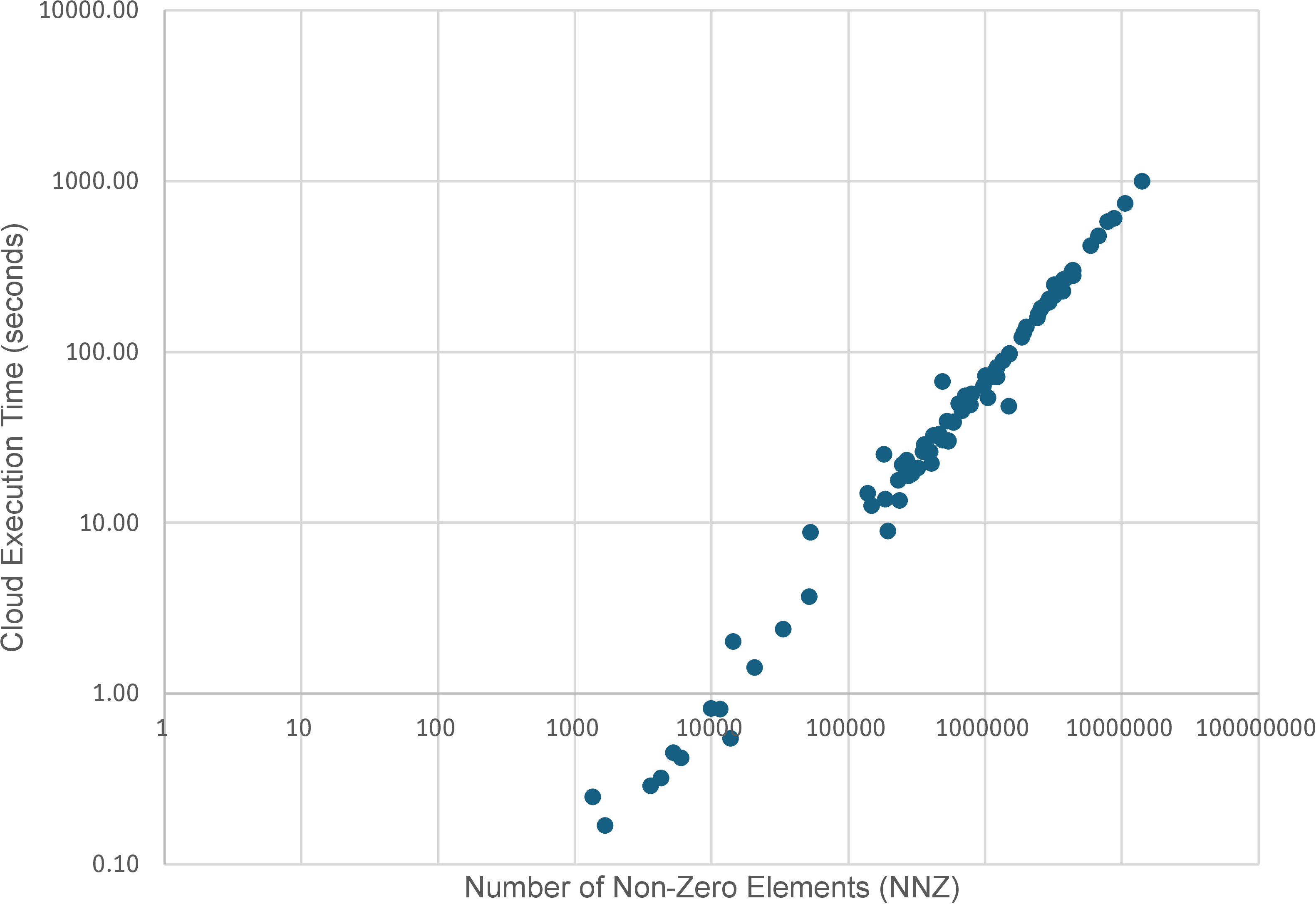}
  \caption{Log-log plot of cloud execution time versus number of non-zero elements (NNZ) for various sparse matrices. The nearly linear trend in log-log space confirms the expected power-law scaling, consistent with the theoretical time complexity of \( O(n \cdot \log C_{\max}) \).}
  \label{fig:loglog_runtime}
\end{figure}

\subsection{Memory Evaluation}
To measure peak memory consumption, we use Python’s built-in \texttt{tracemalloc} module~\citep{python_tracemalloc}, which tracks memory allocations across all phases of execution, including encryption, ciphertext packing, homomorphic computation, and communication. {Because underlying HE libraries allocate memory in native code, we additionally estimate memory for ciphertext-heavy baselines using the measured per-ciphertext size (0.52 MB) and the number of ciphertexts generated.} All reported values represent the maximum snapshot recorded during the computation phase, ensuring a consistent and conservative assessment of the algorithm’s memory footprint.

\begin{table}[t]
  \centering
  \caption{Memory usage comparison between our proposed method and the baseline approaches on sparse matrices under homomorphic encryption. The table reports matrix dimensions (rows, columns, and number of non-zeros), followed by memory consumption (in MB) for four methods: our method, Diag~\citep{halevi2014algorithms}, HEGMM~\citep{gao2024secure}, and HETAL~\citep{lee2023hetal}. The last column shows the memory efficiency gain, computed as the ratio of the best (least) memory usage among three baseline methods to that of our method. Larger values indicate better memory savings. Approximate values are estimated based on the number of ciphertexts, with each ciphertext occupying 0.52 MB as measured in our experiments.
}
\begin{adjustbox}{max width=\linewidth}
\begin{tabular}{c|ccc|cccc|c}

\multicolumn{1}{c|}{\multirow{2}{*}{\textbf{Matrix}}} &
\multicolumn{1}{c}{\multirow{2}{*}{\textbf{Rows}}} &
\multicolumn{1}{c}{\multirow{2}{*}{\textbf{Cols}}} &
\multicolumn{1}{c|}{\multirow{2}{*}{\textbf{NNZ}}} &
\multicolumn{4}{c|}{\textbf{Memory (MB)}} &
\multicolumn{1}{c}{\multirow{2}{*}{\makecell{\textbf{Memory}\\\textbf{Saving (xTimes)}}}} \\
 & & & &
\multicolumn{1}{c}{\textbf{Ours}} &
\multicolumn{1}{c}{\textbf{Diag\citep{halevi2014algorithms}}} &
\multicolumn{1}{c}{\textbf{HEGMM\citep{gao2024secure}}} &
\multicolumn{1}{c|}{\textbf{HETAL\citep{lee2023hetal}}} & \\
\midrule
arc130            & 130   & 130    & 1354     & \bfseries 1.33   & 317.16    & 8.08      & 8.06       & 6.06 \\
stat96v5          & 2307 & 75779 & 233921   & \bfseries 1509.12 & 4916.19  & 29654.11 & 28938.01  & 3.26 \\
M80PI\_n1         & 4028 & 4028  & 9927     & \bfseries 140.42 & 4662.45  & 2541.23  & 2521.64   & 17.96 \\
mycielskian13     & 6143 & 6143  & 1227742 & \bfseries 390.33 & 7059.54  & 5856.08  & 6037.39   & 15.00 \\
nemeth24          & 9506 & 9506  & 1506550 & \bfseries 813.70 & 10924.30 & 14957.53 & 14914.10  & 13.43 \\
msc10848          & 10848& 10848 & 1229776 & \bfseries 1014.04 & 7051.20 & 19175.89 & {\(\approx 15028.00\)} & 6.95 \\
as-caida          & 26475& 26475 & 53381    & \bfseries 7645.86 & 30425.07 & 113725.61& 114161.60 & 3.98 \\
ship\_001         & 34920& 34920 & 3896496 & \bfseries 9619.61 & 40130.06 & \multicolumn{1}{c}{\(\approx 155304.24\)} & {\(\approx 155304.24\)} & 4.17 \\
cca               & 49152& 49152 & 139264   & \bfseries 18636.32 & 56485.48 & \multicolumn{1}{c}{\(\approx 307107.84\)} & {\(\approx 307107.84\)} & 3.03 \\
p2p-Gnutella31    & 62586& 62586 & 147892   & \bfseries 30143.85 & 71923.83 & \multicolumn{1}{c}{\(\approx 497880.24\)} & {\(\approx 497880.24\)} & 2.39 \\

\end{tabular}
\end{adjustbox}
  \label{tab:memo}%
\end{table}%

Table~\ref{tab:memo} presents a memory usage comparison between our proposed method and three baseline approaches for performing sparse matrix operations under homomorphic encryption. The evaluated matrices vary widely in size and sparsity, ranging from small instances like \texttt{arc130} (130 × 130 with 1,354 non-zeros) to much larger structures such as \texttt{p2p-Gnutella31} (62,586 × 62,586 with 147,892 non-zeros).

Across all cases, our method demonstrates significantly lower memory consumption compared to the baselines, with memory efficiency gains ranging from approximately 2.4× to nearly 18×. Notably, on medium-sized matrices such as \texttt{M80PI\_n1} and \texttt{mycielskian13}, our approach achieves memory savings of 17.96× and 15.00×, respectively. Even for large matrices like \texttt{msc10848} and \texttt{ship\_001}, our method reduces memory usage by over 4× to 7× compared to the most efficient baseline.

For the cases where the baseline memory usage is marked with $\approx$, the values were estimated based on the number of ciphertexts and their size, either because the computation could not be completed due to time limits or because it relied on the system's internal memory optimization strategies (e.g., swapping or caching). Even under these favorable assumptions for the baseline approaches, our method still achieves substantial reductions in memory consumption, particularly on high-dimensional matrices with moderate sparsity.

These improvements stem from two main design strengths: (1) our compact sparse ciphertext structure (CSSC), which reduces redundancy in ciphertext representation, and (2) a chunking strategy that minimizes the number of ciphertexts required to represent non-zero elements. In contrast, baseline approaches are burdened by either dense ciphertext packing or inefficient sparse encoding, leading to inflated memory usage.

Overall, the results demonstrate that our method offers not only computational efficiency but also considerable memory savings, making it a practical solution for encrypted sparse computations where storage is a critical constraint. More experiment results are shown in Table~\ref{tab:expapdx} in \ref{apx:exp}.

\subsection{Communication Overhead}
As part of the proposed algorithm, communication occurs between a client (A or B) and the cloud server (D), as well as directly between the two clients. Table~\ref{tab:communication_cost} summarizes the communication overhead associated with each dataset, measured in megabytes (MB). The column ``A $\to$ D / B $\to$ D" represents the amount of data transmitted from each client to the cloud, while the column ``A $\to$ B" indicates the amount of data exchanged directly between the clients. These values reflect the communication overhead involved during the execution of the algorithm and provide a reference for understanding the data transfer requirements across different datasets.

\begin{table}[ht]
\centering
\caption{Communication Overhead among Clients and Cloud. \textbf{A} and \textbf{B} represent clients, and \textbf{Cloud} represents the cloud server. A $\to$ Cloud / B $\to$ Cloud denotes the communication overhead from each client to the cloud. A $\to$ B denotes the communication overhead directly between Client A and Client B.}
\small
\label{tab:communication_cost}
\begin{tabular}{c|c|c}
\textbf{Matrix} & \textbf{A $\to$ Cloud / B $\to$ Cloud (MB)} & \textbf{A $\to$ B (Bytes)} \\
\hline
arc130             & 1.04   & 88  \\
stat96v5           & 53.04  & 920  \\
M80PI\_n1          & 2.08   & 88  \\
mycielskian13      & 183.04 & 3184  \\
nemeth24           & 197.60 & 3560  \\
msc10848           & 176.80 & 3240  \\
as-caida           & 17.16  & 776  \\
ship\_001          & 520.52 & 9240 \\
cca                & 17.68  & 1056  \\
p2p-Gnutella31     & 21.32  & 760  \\

\end{tabular}
\end{table}

\subsection{Evaluation Against the Plaintext-Ciphertext Matrix-Vector Multiplication Baseline Approach}
As noted in Section~\ref{sec:expbase}, Rhombus~\citep{he2024rhombus} follows a plaintext matrix times ciphertext vector design and processes one plaintext row at a time. Table \ref{tab:rhombus} shows that our method runs substantially faster on all datasets. For example, on \texttt{stat96v5} the runtime decreases from 1683.25 seconds to 18.76 seconds, and on \texttt{arc130} from 6.40 seconds to 0.27 seconds. The primary reason for this improvement is the large difference in the number of ciphertexts. Rhombus must perform one encrypted operation (pt*ct) per matrix row, which produces thousands of ciphertext related operations and many rotation steps. In contrast, the CSSC structure merges many rows into a small number of ciphertext chunks, so the cloud performs far fewer multiplications and rotations.

The memory usage follow the same pattern. Although Rhombus uses plaintext to represent matrix rows, the sheer number of these rows forces it to keep many plaintext vectors and intermediate results in memory. This leads to a high peak memory cost, as seen for matrices such as \texttt{msc10848} and \texttt{p2p-Gnutella31}. Our method uses a compact ciphertext representation that preserves sparsity and keeps only a limited number of ciphertext chunks during evaluation. As a result, our peak memory usage is consistently {comparable to or lower than Rhombus, and substantially lower on the largest instances due to fewer ciphertexts and intermediates.}

Overall, even though Rhombus operates under a more favorable trust model, our method shows significantly better time and memory performance because CSSC reduces the number of ciphertexts and therefore the number of homomorphic operations required.

\begin{table}[t]
\centering
\small
\caption{Comparison of our method with Rhombus: runtime (seconds) and memory (MB). 
Our approach performs ciphertext-ciphertext matrix-vector multiplication, while Rhombus performs plaintext matrix ciphertext vector multiplication.} 
\label{tab:rhombus}
\begin{adjustbox}{max width=\linewidth}
\begin{tabular}{c|cc|cc}

\multicolumn{1}{c|}{\multirow{2}{*}{\textbf{Matrix}}} & \multicolumn{2}{c|}{\textbf{Time (s)}} & \multicolumn{2}{c}{\textbf{Memory (MB)}} \\
 & \textbf{Ours} & \textbf{Rhombus} & \textbf{Ours} & \textbf{Rhombus} \\
\midrule
arc130           & 0.27   & 6.40      & 1.33     & 143.44 \\
stat96v5         & 18.76  & 1683.25   & 1509.15  & 1479.00 \\
M80PI\_n1        & 0.83   & 278.98    & 140.44   & 269.12 \\
mycielskian13    & 69.35  & 426.17    & 390.47   & 433.56 \\
nemeth24         & 97.72  & 1312.29   & 813.85   & 837.97 \\
msc10848         & 81.28  & 1518.11   & 1014.16  & 1044.75 \\
as-caida         & 8.83   & 9091.51   & 7645.84  & 7658.15 \\
ship\_001        & 261.40 & 11846.40  & 9619.73  & 9462.98 \\
cca              & 14.85  & 14303.09  & 18636.36 & 18577.58 \\
p2p-Gnutella31   & 12.58  & 37186.32  & 30143.78 & 30028.66 \\

\end{tabular}
\end{adjustbox}
\end{table}

\section{Conclusions, Future Work, and Limitations}

We proposed the first framework, Compressed Sparse Sorted Column (CSSC) format, for sparse matrix–vector multiplication (SpMV) under homomorphic encryption (HE), addressing limitations of prior dense-focused approaches. 
In the extensive experiments with real-world matrices, our method achieves up to five orders of magnitude speedup and substantial memory reduction over state-of-the-art baselines. These gains arise from structure-aware encoding and minimized homomorphic operation depth, allowing deployment under practical HE parameters. 
Overall, our work provides a foundation for scalable and efficient encrypted computation in cloud and privacy-sensitive environments.

There are several extensions remaining open.
First, our current setting targets single-party encrypted computation; extending the design to multi-party or federated environments would broaden applicability and introduce new challenges in secure aggregation. Secondly, while BFV is used in this work, evaluating alternative schemes (\eg, CKKS for approximate arithmetic or FHEW~\cite{DucasMicciancio2015FHEW}/TFHE~\cite{Chillotti2020TFHE} for faster bootstrapping) may offer better trade-offs for specific sparse workloads. Thirdly, CSSC currently assumes a static sparsity pattern. Applications with evolving sparsity would benefit from incremental update mechanisms that avoid full recompression. 
Finally, although our experiments cover diverse sparse matrices, further domain-specific evaluations (\eg, genomics, recommendations, encrypted graph data) would strengthen generality. In short, while the proposed framework provides an efficient foundation for encrypted sparse computation, extending it to dynamic, distributed, and multi-scheme settings represents promising future work.

\section{Acknowledgement}
This work was supported in part by the National Science Foundation of the United States of America (Grant No. 2321572).

\bibliographystyle{elsarticle-num}
\bibliography{ref}

\appendix
\section{Complete Experimental Results}
\label{apx:exp}
\small
\begin{xltabular}{\linewidth}{l|S[table-format=6]|S[table-format=6]|S[table-format=8]|S[table-format=4.2]|S[table-format=6.2]}
\caption{Complete experimental results for all evaluated matrices, including size (rows and columns), number of non-zero elements (NNZ), execution time (in seconds), and memory usage (in megabytes).} 
\label{tab:expapdx} \\
\toprule
\textbf{Matrix} & \textbf{Rows} & \textbf{Cols} & \textbf{NNZ} & \textbf{Time (s)} & \textbf{Mem (MB)}\\
\midrule
\endfirsthead
\toprule
\textbf{Matrix} & \textbf{Rows} & \textbf{Cols} & \textbf{NNZ} & \textbf{Time (s)} & \textbf{Mem (MB)}\\
\midrule
\endhead
\midrule \multicolumn{6}{r}{\textit{Continued on next page}}\\ \endfoot
\bottomrule \endlastfoot

arc130 & 130 & 130 & 1354 & 0.25 & 1.33 \\
494\_bus & 494 & 494 & 1669 & 0.17 & 3.08 \\
fs\_541\_4 & 541 & 541 & 4273 & 0.32 & 3.45 \\
jpwh\_991 & 991 & 991 & 6027 & 0.42 & 9.05 \\
bcspwr06 & 1454 & 1454 & 5300 & 0.45 & 18.81 \\
watt\_2 & 1856 & 1856 & 11550 & 0.81 & 30.48 \\
CAG\_mat1916 & 1916 & 1916 & 195985 & 9.00 & 44.23 \\
stat96v5 & 2307 & 75779 & 233921 & 17.78 & 1509.12 \\
heart2 & 2339 & 2339 & 680341 & 45.29 & 95.92 \\
heart3 & 2339 & 2339 & 680341 & 45.56 & 95.92 \\
yeast & 2361 & 2361 & 13828 & 0.55 & 39.08 \\
qc2534 & 2534 & 2534 & 463360 & 33.16 & 139.33 \\
reorientation\_4 & 2717 & 2717 & 33630 & 2.39 & 65.08 \\
cegb2802 & 2802 & 2802 & 277362 & 18.79 & 82.63 \\
cegb2919 & 2919 & 2919 & 321543 & 20.98 & 91.51 \\
mycielskian12 & 3071 & 3071 & 407200 & 22.38 & 103.54 \\
psmigr\_2 & 3140 & 3140 & 540022 & 30.38 & 118.41 \\
psmigr\_1 & 3140 & 3140 & 543160 & 30.32 & 118.66 \\
psmigr\_3 & 3140 & 3140 & 543160 & 30.08 & 118.66 \\
stat96v4 & 3173 & 63076 & 491336 & 30.43 & 1735.46 \\
raefsky1 & 3242 & 3242 & 293409 & 19.38 & 104.52 \\
raefsky2 & 3242 & 3242 & 293551 & 19.56 & 104.53 \\
bayer05 & 3268 & 3268 & 20712 & 1.42 & 92.87 \\
bcsstm21 & 3600 & 3600 & 3600 & 0.29 & 112.13 \\
nemsemm1 & 3945 & 75352 & 1053986 & 54.22 & 2588.50 \\
M80PI\_n1 & 4028 & 4028 & 9927 & 0.82 & 140.42 \\
ca-GrQc & 5242 & 5242 & 14496 & 2.03 & 232.05 \\
bas1lp & 5411 & 9825 & 587775 & 38.79 & 465.46 \\
stat96v1 & 5995 & 197472 & 588798 & 38.90 & 9824.81 \\
mycielskian13 & 6143 & 6143 & 1227742 & 71.76 & 390.33 \\
EternityII\_A & 7362 & 150638 & 782087 & 49.16 & 9077.57 \\
TSC\_OPF\_1047 & 8140 & 8140 & 2012833 & 140.15 & 699.91 \\
nemeth20 & 9506 & 9506 & 971870 & 63.10 & 782.24 \\
nemeth21 & 9506 & 9506 & 1173746 & 76.71 & 793.48 \\
nemeth22 & 9506 & 9506 & 1358832 & 89.10 & 804.21 \\
nemeth24 & 9506 & 9506 & 1506550 & 97.15 & 813.70 \\
nemeth23 & 9506 & 9506 & 1506810 & 97.60 & 813.47 \\
nemeth25 & 9506 & 9506 & 1511758 & 97.65 & 813.90 \\
nemeth26 & 9506 & 9506 & 1511760 & 97.60 & 813.90 \\
ca-HepTh & 9877 & 9877 & 51971 & 3.71 & 788.41 \\
EternityII\_Etilde & 10054 & 204304 & 1170516 & 71.75 & 16546.90 \\
msc10848 & 10848 & 10848 & 1229776 & 81.58 & 1014.04 \\
EternityII\_E & 11077 & 262144 & 1503732 & 48.01 & 23231.46 \\
ca-HepPh & 12008 & 12008 & 237010 & 13.55 & 1166.96 \\
mycielskian14 & 12287 & 12287 & 3695512 & 227.71 & 1441.94 \\
mycielskian14 & 12287 & 12287 & 3695512 & 227.67 & 1441.94 \\
appu & 14000 & 14000 & 1853104 & 122.15 & 1648.41 \\
opt1 & 15449 & 15449 & 1930655 & 129.11 & 1982.25 \\
ramage02 & 16830 & 16830 & 2866352 & 194.58 & 2373.74 \\
pkustk07 & 16860 & 16860 & 2418804 & 159.01 & 2360.05 \\
TSOPF\_RS\_b678\_c1 & 18696 & 18696 & 4396289 & 300.06 & 2990.01 \\
ca-AstroPh & 18772 & 18772 & 396160 & 26.18 & 2790.39 \\
tsyl201 & 20685 & 20685 & 2454957 & 166.05 & 3460.68 \\
pkustk08 & 22209 & 22209 & 3226671 & 213.64 & 4007.60 \\
ca-CondMat & 23133 & 23133 & 186936 & 13.77 & 4187.52 \\
TSOPF\_RS\_b2052\_c1 & 25626 & 25626 & 6761100 & 479.62 & 5465.63 \\
TSOPF\_RS\_b2052\_c1 & 25626 & 25626 & 6761100 & 476.57 & 5465.63 \\
smt & 25710 & 25710 & 3749582 & 265.11 & 5318.58 \\
as-caida & 26475 & 26475 & 53381 & 8.77 & 7645.86 \\
cit-HepTh & 27770 & 35963 & 352807 & 26.16 & 6015.74 \\
TSOPF\_RS\_b300\_c2 & 28338 & 28338 & 2943887 & 196.25 & 6381.86 \\
TSOPF\_FS\_b300 & 29214 & 29214 & 4400122 & 279.48 & 6875.61 \\
TSOPF\_FS\_b300\_c1 & 29214 & 29214 & 4400122 & 279.81 & 6875.61 \\
thread & 29736 & 29736 & 4444880 & 299.70 & 7063.60 \\
cit-HepPh & 34546 & 42339 & 421578 & 32.36 & 9267.22 \\
ship\_001 & 34920 & 34920 & 3896496 & 268.77 & 9619.61 \\
TSOPF\_RS\_b678\_c2 & 35696 & 35696 & 8781949 & 603.99 & 10263.74 \\
pdb1HYS & 36417 & 36417 & 4344765 & 294.52 & 10444.81 \\
email-Enron & 36692 & 36692 & 183831 & 25.26 & 10443.22 \\
windtunnel\_evap3d & 40816 & 40816 & 803978 & 56.95 & 12907.60 \\
TSOPF\_RS\_b300\_c3 & 42138 & 42138 & 4413449 & 297.18 & 13908.05 \\
pkustk06 & 43164 & 43164 & 2571768 & 176.94 & 14494.90 \\
mark3jac100 & 45769 & 45769 & 268563 & 23.35 & 16178.62 \\
mosfet2 & 46994 & 46994 & 1013062 & 72.58 & 17080.55 \\
c-65 & 48066 & 48066 & 360428 & 28.83 & 17839.21 \\
cca & 49152 & 49152 & 139264 & 14.89 & 18636.32 \\
rajat26 & 51032 & 51032 & 247528 & 21.81 & 20086.69 \\
ct20stif & 52329 & 52329 & 2600295 & 180.54 & 21207.51 \\
crankseg\_1 & 52804 & 52804 & 10614210 & 741.84 & 21923.13 \\
g7jac180sc & 53370 & 53370 & 641290 & 49.71 & 21983.89 \\
srb1 & 54924 & 54924 & 2962152 & 203.89 & 23349.09 \\
c-67b & 57975 & 57975 & 530583 & 39.46 & 25905.52 \\
g7jac200sc & 59310 & 59310 & 717620 & 55.48 & 27117.89 \\
Ga3As3H12 & 61349 & 61349 & 5970947 & 418.09 & 29332.46 \\
p2p-Gnutella31 & 62586 & 62586 & 147892 & 12.62 & 30143.85 \\
crankseg\_2 & 63838 & 63838 & 14148858 & 1002.04 & 31918.81 \\
Goodwin\_095 & 100037 & 100037 & 3226066 & 247.99 & 76875.57 \\
lung2 & 109460 & 109460 & 492564 & 67.21 & 91871.54 \\
Ge87H76 & 112985 & 112985 & 7892195 & 581.36 & 98188.38 \\
\end{xltabular}

\section{Psudo code for algorithm}
\label{apx:alg}

Algorithm~\ref{alg:csrtocssc} converts a sparse matrix $M \in \mathbb{R}^{r \times c}$ from the CSR
representation $\bigl(V^{\mathrm{csr}}, CI^{\mathrm{csr}}, RP^{\mathrm{csr}}\bigr)$ into the proposed CSSC
format $(V, CI, CP, RM)$ by reorganizing non-zeros into a left-aligned, column-major layout while
preserving each non-zero value and its original column index. It first computes the number of non-zeros
in each row and the maximum row length:
\[
\text{nnzRow}[i] = RP^{\mathrm{csr}}[i+1] - RP^{\mathrm{csr}}[i], \qquad
L = \max_{0 \le i < r} (\text{nnzRow}[i]), \qquad
\text{nnz} = RP^{\mathrm{csr}}[r].
\]
Rows are then permuted in descending order of $\text{nnzRow}[i]$ to form the row map
\[
RM = \operatorname{argsort}\!\bigl(\text{nnzRow}\bigr),
\]
which places longer rows first and improves the regularity of left alignment. After allocating
$V \in \mathbb{R}^{\text{nnz}}$, $CI \in \mathbb{Z}^{\text{nnz}}$, and $CP \in \mathbb{Z}^{L+1}$, the conversion
proceeds by scanning the left-aligned rows in a column-major fashion. For each within-row position
$j = 0,1,\dots,L-1$ and each permuted row index $i = RM[p]$ ($p = 0,1,\dots,r-1$), if $j < \text{nnzRow}[i]$,
the CSR index of the $j$-th non-zero in row $i$ is
\[
k = RP^{\mathrm{csr}}[i] + j,
\]
and the entry is appended to the CSSC arrays:
\[
V[w] = V^{\mathrm{csr}}[k], \qquad
CI[w] = CI^{\mathrm{csr}}[k], \qquad
w \leftarrow w + 1.
\]
The pointer array stores the starting offset of each aligned group via $CP[j]=w$ (before processing
position $j$), and the final pointer satisfies
\[
CP[L] = w = \text{nnz}.
\]
Consequently, each segment $V[CP[j]:CP[j+1])$ (and the corresponding $CI$ segment) contains all
$j$-th non-zeros across rows in the permuted order, yielding the CSSC layout.

\begin{algorithm}[htbp]
\small
\caption{Conversion from CSR to CSSC Format}
\label{alg:csrtocssc}
\begin{algorithmic}[1]
\REQUIRE CSR arrays $V^{\mathrm{csr}}$, $CI^{\mathrm{csr}}$, $RP^{\mathrm{csr}}$ for $M \in \mathbb{R}^{r \times c}$
\ENSURE CSSC arrays $(V, CI, CP, RM)$

\text{/*\textit{Compute number of non-zeros per row}*/}
\FOR{$i = 0$ \TO $r-1$}
    \STATE $\text{nnzRow}[i] \gets RP^{\mathrm{csr}}[i+1] - RP^{\mathrm{csr}}[i]$
\ENDFOR
\STATE $L \gets \max_i \text{nnzRow}[i]$
\STATE $\text{nnz} \gets RP^{\mathrm{csr}}[r]$
\text{/*\textit{Row permutation: sort rows by decreasing nnz}*/}
\STATE $RM \gets \text{argsort of row indices in descending order of } \text{nnzRow}$

\text{/*\textit{Allocate CSSC arrays}*/}
\STATE allocate $V[0..\text{nnz}-1]$, $CI[0..\text{nnz}-1]$, $CP[0..L]$
\STATE $w \gets 0$

\text{/*\textit{Column-major traversal of left-aligned rows}*/}
\FOR{$j = 0$ \TO $L-1$}
    \STATE $CP[j] \gets w$
    \FOR{$p = 0$ \TO $r-1$}
        \STATE $i \gets RM[p]$ \text{/*\textit{original row index}*/}
        \STATE $k_0 \gets RP^{\mathrm{csr}}[i]$
        \STATE $\ell \gets \text{nnzRow}[i]$
        \IF{$j < \ell$}
            \STATE $k \gets k_0 + j$ \text{/*\textit{index of the $j$-th non-zero in row $i$ (CSR order)}*/}
            \STATE $V[w] \gets V^{\mathrm{csr}}[k]$
            \STATE $CI[w] \gets CI^{\mathrm{csr}}[k]$ \text{/*\textit{original column index in $M$}*/}
            \STATE $w \gets w + 1$
        \ENDIF
    \ENDFOR
\ENDFOR
\STATE $CP[L] \gets w$

\end{algorithmic}
\end{algorithm}

\end{document}